\documentclass[twocolumn]{aastex63}

\usepackage{tabularx}

\received{March 09, 2021}
\revised{April 09, 2021}
\accepted{April 16, 2021}
\submitjournal{ApJ}

\shorttitle{The $\alpha$-element abundances of field RR Lyrae variables}
\shortauthors{Crestani et al.}

\graphicspath{{./}{}}

\begin{document}

\title{On the Use of Field RR Lyrae as Galactic Probes. III.
The $\alpha$-element abundances 
\footnote{Based on observations obtained with the du Pont telescope at Las Campanas Observatory, operated by Carnegie Institution for Science.
Based in part on data collected at Subaru Telescope, which is operated by the National Astronomical Observatory of Japan.
Based partly on data obtained with the STELLA robotic telescopes in Tenerife, an AIP facility jointly operated by AIP and IAC.
Some of the observations reported in this paper were obtained with the Southern African Large Telescope (SALT).
Based on observations made with the Italian Telescopio Nazionale Galileo (TNG) operated on the island of La Palma by the Fundación Galileo 
Galilei of the INAF (Istituto Nazionale di Astrofisica) at the Spanish Observatorio del Roque de los Muchachos of the Instituto de Astrofisica de Canarias.
Based on observations collected at the European Organisation for Astronomical Research in the Southern Hemisphere.
}
}

\correspondingauthor{J. Crestani}
\email{juliana.crestani@uniroma1.it}

\author{J. Crestani}
\affiliation{Dipartimento di Fisica, Universit\`a di Roma Tor Vergata, via della Ricerca Scientifica 1, 00133 Roma, Italy}
\affiliation{INAF -- Osservatorio Astronomico di Roma, via Frascati 33, 00078 Monte Porzio Catone, Italy}
\affiliation{Departamento de Astronomia, Universidade Federal do Rio Grande do Sul, Av. Bento Gon\c{c}alves 6500, Porto Alegre 91501-970, Brazil}


\author{V. F. Braga}
\affiliation{INAF -- Osservatorio Astronomico di Roma, via Frascati 33, 00078 Monte Porzio Catone, Italy}
\affiliation{Space Science Data Center -- ASI, via del Politecnico snc, 00133 Roma, Italy}

\author{M. Fabrizio}
\affiliation{INAF -- Osservatorio Astronomico di Roma, via Frascati 33, 00078 Monte Porzio Catone, Italy}
\affiliation{Space Science Data Center -- ASI, via del Politecnico snc, 00133 Roma, Italy}

\author{G. Bono}
\affiliation{Dipartimento di Fisica, Universit\`a di Roma Tor Vergata, via della Ricerca Scientifica 1, 00133 Roma, Italy}
\affiliation{INAF -- Osservatorio Astronomico di Roma, via Frascati 33, 00078 Monte Porzio Catone, Italy}

\author{C. Sneden}
\affiliation{Department of Astronomy and McDonald Observatory, The University of Texas, Austin, TX 78712, USA}

\author{G. Preston}
\affiliation{The Observatories of the Carnegie Institution for Science, 813 Santa Barbara St., Pasadena, CA 91101, USA}

\author{I. Ferraro}
\affiliation{INAF -- Osservatorio Astronomico di Roma, via Frascati 33, 00078 Monte Porzio Catone, Italy}

\author{G. Iannicola}
\affiliation{INAF -- Osservatorio Astronomico di Roma, via Frascati 33, 00078 Monte Porzio Catone, Italy}

\author{M. Nonino}
\affiliation{INAF -- Osservatorio Astronomico di Trieste, Via G. B. Tiepolo 11, 34143 Trieste, Italy}

\author{G. Fiorentino}
\affiliation{INAF -- Osservatorio Astronomico di Roma, via Frascati 33, 00078 Monte Porzio Catone, Italy}

\author{F. Th\'evenin}
\affiliation{Universit\'e de Nice Sophia-antipolis, CNRS, Observatoire de la C\^ote d'Azur, Laboratoire Lagrange, BP 4229, F-06304 Nice, France}

\author{B. Lemasle}
\affiliation{Astronomisches Rechen-Institut, Zentrum f\"ur Astronomie der Universit\"at Heidelberg, M\"onchhofstr. 12-14, D-69120
Heidelberg, Germany}

\author{Z. Prudil}
\affiliation{Astronomisches Rechen-Institut, Zentrum f\"ur Astronomie der Universit\"at Heidelberg, M\"onchhofstr. 12-14, D-69120
Heidelberg, Germany}


\author{A. Alves-Brito}
\affiliation{Departamento de Astronomia, Universidade Federal do Rio Grande do Sul, Av. Bento Gon\c{c}alves 6500, Porto Alegre 91501-970, Brazil}

\author{G. Altavilla}
\affiliation{INAF -- Osservatorio Astronomico di Roma, via Frascati 33, 00078 Monte Porzio Catone, Italy}
\affiliation{Space Science Data Center -- ASI, via del Politecnico snc, 00133 Roma, Italy}

\author{B. Chaboyer}
\affiliation{Department of Physics and Astronomy, Dartmouth College, Hanover, NH 03784, USA}

\author{M. Dall'Ora}
\affiliation{INAF -- Osservatorio Astronomico di Capodimonte, Salita Moiariello 16, 80131 Napoli, Italy}

\author{V. D'Orazi}
\affiliation{INAF -- Osservatorio Astronomico di Padova, vicolo dell’Osservatorio 5, 35122, Padova, Italy}
\affiliation{School of Physics and Astronomy, Monash University, Clayton, VIC 3800, Melbourne, Australia}

\author{C. Gilligan}
\affiliation{Department of Physics and Astronomy, Dartmouth College, Hanover, NH 03784, USA}

\author{E. K. Grebel} 
\affiliation{Astronomisches Rechen-Institut, Zentrum f\"ur Astronomie der Universit\"at Heidelberg, M\"onchhofstr. 12-14, D-69120
Heidelberg, Germany}

\author{A. J. Koch-Hansen}
\affiliation{Astronomisches Rechen-Institut, Zentrum f\"ur Astronomie der Universit\"at Heidelberg, M\"onchhofstr. 12-14, D-69120
Heidelberg, Germany}

\author{H. Lala}
\affiliation{Astronomisches Rechen-Institut, Zentrum f\"ur Astronomie der Universit\"at Heidelberg, M\"onchhofstr. 12-14, D-69120
Heidelberg, Germany}

\author{M. Marengo}
\affiliation{Department of Physics and Astronomy, Iowa State University, Ames, IA 50011, USA}

\author{S. Marinoni}
\affiliation{INAF -- Osservatorio Astronomico di Roma, via Frascati 33, 00078 Monte Porzio Catone, Italy}
\affiliation{Space Science Data Center -- ASI, via del Politecnico snc, 00133 Roma, Italy}

\author{P. M. Marrese}
\affiliation{INAF -- Osservatorio Astronomico di Roma, via Frascati 33, 00078 Monte Porzio Catone, Italy}
\affiliation{Space Science Data Center -- ASI, via del Politecnico snc, 00133 Roma, Italy}

\author{C. Mart\'inez-V\'azquez}
\affiliation{Cerro Tololo Inter-American Observatory, NSF's National Optical-Infrared Astronomy Research Laboratory, Casilla
603, La Serena, Chile}

\author{N. Matsunaga}
\affiliation{Department of Astronomy, The University of Tokyo, 7-3-1 Hongo, Bunkyo-ku, Tokyo 113-0033, Japan} 

\author{M. Monelli}
\affiliation{Instituto de Astrof\'isica de Canarias, Calle Via Lactea s/n, E38205 La Laguna, Tenerife, Spain}

\author{J. P. Mullen}
\affiliation{Department of Physics and Astronomy, Iowa State University, Ames, IA 50011, USA}

\author{J. Neeley}
\affiliation{Department of Physics, Florida Atlantic University, 777 Glades Rd, Boca Raton, FL 33431 USA}

\author{R. da Silva}
\affiliation{INAF -- Osservatorio Astronomico di Roma, via Frascati 33, 00078 Monte Porzio Catone, Italy}
\affiliation{Space Science Data Center -- ASI, via del Politecnico snc, 00133 Roma, Italy}

\author{P. B. Stetson}
\affiliation{Herzberg Astronomy and Astrophysics, National Research Council, 5071 West Saanich Road, Victoria, British
Columbia V9E 2E7, Canada}

\author{M. Salaris}
\affiliation{Astrophysics Research Institute, Liverpool John Moores University, IC2, Liverpool Science Park, 146 Brownlow Hill, Liverpool,L3 5RF, United Kingdom}

\author{J. Storm}
\affiliation{Leibniz-Institut f\"ur Astrophysik Potsdam (AIP), An der Sternwarte 16, D-14482 Potsdam, Germany}

\author{E. Valenti}
\affiliation{European Southern Observatory, Karl-Schwarzschild-Str. 2, 85748 Garching bei Munchen, Germany}
\affiliation{Excellence Cluster ORIGINS, Boltzmann\--Stra\ss e 2, D\--85748 Garching bei M\"{u}nchen, Germany}

\author{M. Zoccali}
\affiliation{Instituto de Astrof\'sica, Facultad de F\'sica, Pontificia Universidad Cat\'lica de Chile, Av. Vicu\~na Mackenna 4860, Santiago, Chile}

\begin{abstract}
We provide the largest and most homogeneous sample of $\alpha$-element (Mg, Ca, Ti) and iron abundances for field RR Lyrae (RRLs, 162 variables) by using high-resolution  spectra. The current measurements were complemented with similar abundances available  in the literature for 46 field RRLs brought to our metallicity scale. We ended up with a sample of old (t$\ge$ 10 Gyr), low-mass stellar tracers (208 RRLs: 169 fundamental, 38 first overtone, 1 mixed mode) covering three dex in iron abundance (-3.00$\le$[Fe/H]$\le$0.24). We found that field RRLs are $\sim$0.3 dex more $\alpha$-poor than typical Halo tracers in the metal-rich regime, ([Fe/H]$\ge$-1.2) while in the metal-poor regime ([Fe/H]$\le$-2.2) they seem to be on  average $\sim$0.1 dex more $\alpha$-enhanced. This is the first time that the depletion in $\alpha$-elements for solar iron abundances is detected on the basis of a large, homogeneous and coeval sample of old stellar tracers. Interestingly, we also detected a close similarity in the [$\alpha$/Fe] trend between $\alpha$-poor, metal-rich RRLs and red giants (RGs) in the Sagittarius dwarf galaxy as well as between $\alpha$-enhanced, metal-poor RRLs and RGs in ultra faint dwarf galaxies. These results are supported by similar elemental abundances for 46  field Horizontal Branch (HB) stars. These stars share  with RRLs the same evolutionary phase and the same progenitors. This evidence further supports the key role that old  stellar tracers play in constraining the early chemical enrichment of the Halo and, in particular, in investigating the impact that dwarf galaxies have had in the mass assembly of the Galaxy.
\end{abstract}


\section{Introduction} \label{sec:intro}
The chemical abundances of stellar atmospheres preserve the signature of the molecular clouds that
formed them. While some of their elements can be altered during 
stellar evolution, such as the LiCNO group and (rarely) some neutron-capture elements, stellar
atmospheres remain the ideal subjects of Galactic archeology. Different chemical species are formed by processes with their 
own mass and time scales and, coupled with the age of the stellar tracers of interest, reveal the chemical 
enrichment history of different components of the Galaxy. 

The even-Z light elements that are multiples of He nuclei are called $\alpha$-elements. In the past, 
it was believed that they were created by the successive capture of He nuclei. 
However, it is now understood that, while several elements are commonly grouped under the banner of 
$\alpha$-elements, not all of them are created equally nor are all of them equally easy to measure 
\citep{1995ApJS..101..181W,2016PASA...33...40M,2019ApJ...870....2C}. 
In particular the noble gasses Ne and Ar cannot be detected, and S is rarely 
measured in optical spectra \citep{2004ARA&A..42..385G}. A similar limitation applies 
to O, indeed O abundances are typically based either on two weak OI forbidden lines (6300, 6363 \AA) 
or on OI triplet lines (7774, 9263 \AA) that are affected by temperature uncertainties and by non-LTE 
effects \citep{1997ARA&A..35..503M}. Like oxygen, the much easier to measure Mg is created in the 
hydrostatic evolution of massive stars and released on supernovae type II (SNe II) events. 

It is affected by the reaction mechanisms known as the "p-process" 
\citep{1997RvMP...69..995W}, i.e., the production of proton-rich nuclei by a proton-capture mechanism. 
The other three species with easily detectable lines are Si, Ca and Ti. Of these, the first two are 
likely mainly produced during SNe II events, being thus considered "explosive"
$\alpha$-elements. 
The third, Ti, has a vast number of absorption lines that can 
be detected on a broad wavelength and metallicity range.
It is sometimes considered as an iron-peak 
element \citep{1995ApJS...98..617T} with possibly multiple formation channels. Indeed, its dominant
isotope is actually $^{\text{48}}_{\text{22}}$Ti, which is not a multiple of an $\alpha$
particle. Yet the trend of titanium with metallicity follows quite well those of other $\alpha$-elements,
and suggests that, regardless of the precise formation channels, these chemical species are formed 
at similar rates in similar astrophysical sites.

Stellar evolution models point to SNe Ia as the result of the thermonuclear explosions triggered by
the binary interaction between an accreting white dwarf and its companion. The presence of a white
dwarf implies a time scale of the order of billions of years. The yields of such explosions carry 
mostly iron, and so an environment enriched mostly by SNe Ia would have a decreasing [$\alpha$/Fe]
ratio as iron abundance increases. Indeed, in the [$\alpha$/Fe] versus [Fe/H] plane, this decrease is
commonly called the "knee", and is associated with the metallicity at which the SNe Ia began to 
dominate the enrichment of the interstellar medium \citep[e.g., ][]{1990ApJ...365..539M}.

The main producers of $\alpha$-elements, however, are the SNe II. They are the result of the core-collapse of 
massive stars ($\gtrsim$8 M$_{\odot}$), with a time scale of the order of one to ten million years. They 
enrich the interstellar medium with both iron and $\alpha$-elements, with the yields of the latter 
increasing as the mass of the SNe II progenitor increases \citep{2006ApJ...653.1145K}. This means that 
the slope of the [$\alpha$/Fe] abundance ratio and, in particular, its spread at fixed iron content can 
provide firm constraints on the variation of the initial mass Function (IMF) as a function of both time 
and environment \citep{2013ApJ...778..149M,2014ApJ...785..102H,2014A&A...572A..88L,2020A&A...641A.127R}. 

Thus, the fine structure of the [$\alpha$/Fe] abundance ratio as a function of iron abundance has been in
the intersection of several theoretical and empirical investigations \citep{1986A&A...154..279M}. 
For over forty years, $\alpha$-elements have 
been known to be enhanced in primarily old and metal-poor
populations such as field Halo stars and globular clusters (GCs)
\citep{1963ApJ...137..280W,1979ApJ...234..964S,1981ApJ...247..869C,1983ApJS...52..241P,1997ARA&A..35..503M,2004AJ....128.1177V,2005AJ....130.2140P,2009A&A...505..139C}.
The current evidence is suggesting a steady decrease in the  [$\alpha$/Fe] abundance ratio 
for iron abundances more metal-rich than [Fe/H]$\approx$-0.7. However, the number of truly old, 
metal-rich stellar tracers is quite limited \citep[see Figs. 10 and 11 in][]{2011A&A...530A..54G}.  
Indeed, it is not clear yet whether old stellar tracers display the same slope in the [$\alpha$/Fe] 
versus [Fe/H] plane as intermediate and young disk stellar populations in approaching solar iron abundance.  

Two major concerns are involved in the selection of the stellar sample to be employed
in the investigation of the chemical enrichment history of the Halo. 
First, although the [$\alpha$/Fe] abundance ratio is a solid diagnostic, stars covering
a broad range in iron abundances with homogeneous and accurate estimates are necessary. 
Second, any discussion of chemical enrichment history must necessarily take into 
account the age of the stellar tracers that are being employed. Individual age 
estimates for field stars require very precise reddening and distance measurements, and therefore samples of 
field stars suffer various degrees of contamination. For field RGs this limitation
becomes even more severe because they originate from progenitors that
cover a broad range in mass. The natural targets for age-related experiments are GCs because 
they have accurate individual age estimates provided by isochrone fitting
\citep[e.g.][]{1998A&A...335..943S,2013ApJ...775..134V}. However, the current spectroscopic 
investigations that are focused on GCs include only a few metal-rich systems 
\citep{2005AJ....130.2140P,2009A&A...505..139C,2009A&A...505..117C,2010ApJ...712L..21C,2011A&A...530A..54G}.

These two concerns can be addressed at once with the use of field RR Lyrae variable
stars. RRLs are solid tracers of old stellar populations, well known to be evolved low mass
stars with ages necessarily greater than 10 Gyr \citep{2019MNRAS.490.4121W,2020A&A...641A..96S}. They can be
identified by the shape, period, and amplitude of their photometric light curves, all of which are reddening- 
and distance-independent. Their classification, coupled with with spectroscopic atmospheric parameters, provides a strongly univocal identification and make any
sample contamination extremely unlikely. The RRLs also are known to cover a broad range in metallicity 
\citep{2012Ap&SS.341...89W,2017ApJ...835..187C,2017ApJ...848...68S,2021ApJ...908...20C}
and far outnumber
GCs. Indeed, while only roughly 180 GCs have been identified in the Galaxy 
\citep{1996AJ....112.1487H,2010arXiv1012.3224H}, the number of 
field RRLs thanks to long term photometric surveys and to {\em Gaia} is at least a thousand 
times larger\footnote{The current number is still severely underestimated because we lack a 
complete census of RRLs in the inner Bulge and beyond.}. This means that the RRLs can trace 
the variation of chemical abundances across the Galactic spheroid with very high spatial resolution.     
The variation in stellar 
mass of RRLs is at most of the order of 30-40\%, i.e. from $\approx$0.60 to $\approx$0.85 $M_\odot$,
therefore, they are only minimally affected by the shape of the IMF. Moreover, the RRLs are also minimally 
affected by the time dependence, because they formed on a 
time interval of the order of two Gyr.     

In the current work, we aim to investigate the impact that stellar age has on the [$\alpha$/Fe] vs [Fe/H]
plane using high resolution, high signal-to-noise ratio (SNR) spectra of 162 field RRLs. Among them, 138 are fundamental mode 
pulsators (RRab), 23 first overtone pulsators (RRc), and one mixed mode pulsator (RRd). This data set is described in 
Sect.~\ref{ch:dataset}. The current homogeneous measurements of $\alpha$-element and iron abundances
were complemented with similar measurements for field RRLs and HB stars available in the literature and
brought to our scale, as described in Sect.~\ref{ch:samples}.
In Sect.~\ref{ch:measurements} we address how the atmospheric parameters and chemical abundances were
computed. Results are shown and discussed in Sect.~\ref{ch:results}. In Sect.~\ref{ch:final_remarks},
we summarize this investigation and briefly outline future perspectives.

\section{Spectroscopic data set} \label{ch:dataset}
We collected a sample of 407 high resolution spectra for 162 field RRLs (138 RRab, 23 RRc, 1 RRd). In order
to obtain a high enough SNR ($\gtrapprox$ 50 per pixel) for chemical abundance analysis, we stacked low SNR spectra of 
the same star acquired at the same phase and with the same spectrograph, as described in more detail further below.
This process resulted in 243 spectra that were analysed individually.Among them, 51 were acquired with the echelle 
spectrograph at du Pont (Las Campanas Observatory), 74 with UVES \citep{2000SPIE.4008..534D} and 
16 with X-shooter \citep{2011A&A...536A.105V} at VLT (ESO, Cerro Paranal Observatory), 18 with HARPS \citep{2003Msngr.114...20M}
 at the 3.6m telescope and two with FEROS \citep{1999Msngr..95....8K} at the 2.2m MPG telescope 
(ESO, La Silla Observatory), five with HARPS-N \citep{2012SPIE.8446E..1VC} at the Telescopio Nazionale Galileo (Roque de Los 
Muchachos Observatory), 47 with HRS \citep{2014SPIE.9147E..6TC} at SALT (South African Astronomical Observatory), 
28 with the HDS \citep{2002PASJ...54..855N} at Subaru (National Astronomical Observatory of Japan), and two with the 
echelle spectrograph \citep{2012SPIE.8451E..0KW} at STELLA (Iza\~na Observatory). 

\begin{figure}
\includegraphics[width=\columnwidth]{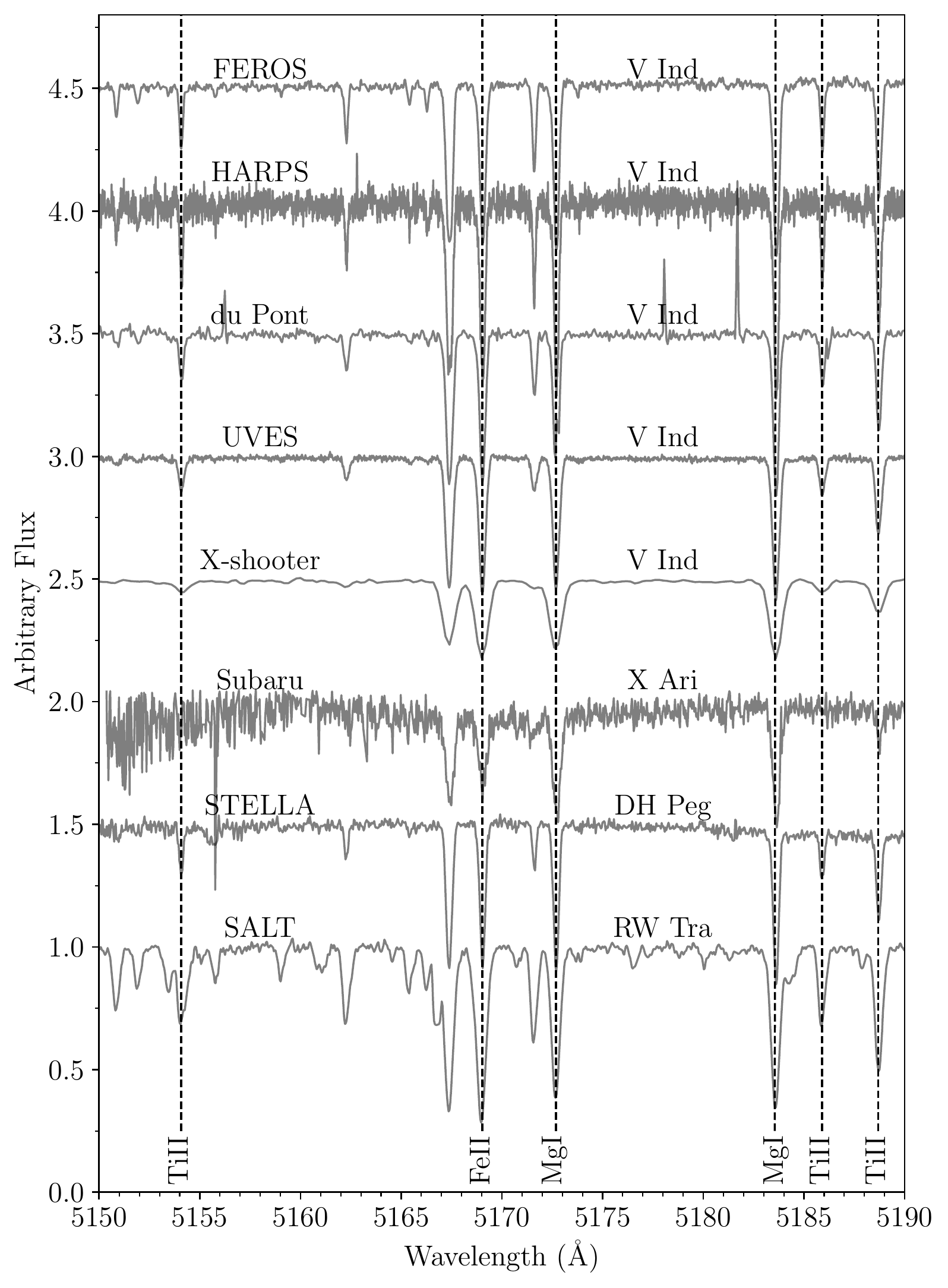}
\caption{Representative high resolution spectra for all spectrographs used in this work. 
The top five spectra are of V Ind ([Fe/H] = -1.63$\pm$± 0.03, RRab) at the same pulsation phase. They are followed by 
random phase spectra for X Ari ([Fe/H] = -2.59$\pm$0.05, RRab), DH Peg ([Fe/H] = -1.37$\pm$0.05, RRc), and RW Tra 
([Fe/H] = 0.13$\pm$0.06, RRab). 
The dashed lines indicate the iron and $\alpha$-element absorption lines in this wavelength region.
\label{fig:spectra_examples}}
\end{figure}

Representative spectra for each of these spectrographs are shown in Fig.~\ref{fig:spectra_examples}. 
Their typical wavelength ranges, resolutions and SNR are listed in Tab.~\ref{tab:spectrographs}.

\begin{deluxetable}{lllrr} 
\tablecaption{Typical characteristics of each instrument used in this work.
\label{tab:spectrographs}}
\tablewidth{\columnwidth}
\tablehead{
\colhead{ Spectrograph}    & \colhead{Telescope} & \colhead{Wavelength}& \colhead{Resolution} & \colhead{SNR} \\ [-0.2cm]
\colhead{ }               & \colhead{ }         & \colhead{Range}      & \colhead{ }          & \colhead{ } \\
\colhead{ }               & \colhead{ }         & \colhead{(\AA)}      & \colhead{ }          & \colhead{ } 
}
\startdata
echelle     & du Pont   & 3700 -- 9100 &  27,000              & 70 \\
UVES        & VLT       & 3000 -- 6800 &  35,000 -- 107,000   & 76 \\
X-shooter   & VLT       & 3000 --10200 &  18,400              & 86 \\
HARPS       & 3.6m      & 3700 -- 6900 &  80,000 -- 115,000   & 45 \\
FEROS       & 2.2m MPG  & 3500 -- 9200 &  48,000              & 53 \\
HARPS-N     & TNG       & 3900 -- 6900 & 115,000              & 65 \\
HRS         & SALT      & 3900 -- 8800 &  40,000              & 61 \\
HDS         & Subaru    & 5060 -- 7840 &  60,000              & 95 \\
echelle     & STELLA    & 3860 -- 8820 &  55,000              & 74 \\
\enddata
\tablecomments{
The wavelength ranges and resolutions are approximate. Different instrumental configurations result in 
different values, including wavelength coverage gaps. The archival data for UVES displayed a significant 
variety of configurations. Only the most representative values are shown.
}
\end{deluxetable}

Continuum normalization and Doppler-shift corrections were made using the National Optical Astronomy Observatory
libraries for {\textsc{IRAF}}\footnote{The legacy code is now maintained by the community on GitHub at 
\url{https://iraf-community.github.io/}} \citep[Image Reduction and Analysis Facility,][]{1993ASPC...52..173T}. 
Further information about the sample selection and radial velocity studies for the spectra analysed in this work
can be found in \citet{2019ApJ...882..169F,2020ApJ...896L..15B}, and \citet{2021ApJ...908...20C}.

The stacking of spectra was performed after these steps. We made an initial selection based on phase, followed by a 
visual inspection. This ensured that all the spectra to be stacked displayed similar line depths and introduced no 
artifacts in the final stacked spectrum. Of the 243 final spectra that we analysed, 178 were collected with high SNR
and did not require stacking, 26 were the result of the stacking of two spectra, and 39 of three to seven spectra.

\section{Spectroscopic samples} \label{ch:samples}
The 162 RRLs described above form the This Work (TW-RRL) sample. Previous high resolution metallicity measurements are
available in the literature for 47 of these stars. They were used to transform iron and $\alpha$-element 
abundances into a homogeneous abundance scale. This supplied us with 69 measurements for 46 stars made by nine 
previous works. Note that the measurements taken from
\citet{2010AJ....140.1694F,2011ApJS..197...29F,2017ApJ...835..187C} and \citet{2017ApJ...848...68S} are natively in our scale and require no shifts.
Indeed, a comparison between 23 RRLs measured both in those works and in the present work resulted in absolute differences smaller than 0.10 dex for all abundances. Note that the investigation from \citet{2010AJ....140.1694F} was focused on non-variable HB stars for which we did not perform a reanalysis, but they used the same line list, instrument, and methodology as \citet{2011ApJS..197...29F}.
Once all abundances of interest were brought to our scale, multiple measurements for the same star were averaged. 
This allowed us to form the Literature RRL (Lit-RRL) sample, with 46 stars. We will refer to the TW-RRL and Lit-RRL 
samples together as the RRL sample. Its basic characteristics are shown in Tab.~\ref{tab:stars}. The individual measurements 
for the literature stars both in their native scale and 
in our scale are shown in Tab.~\ref{tab:literature_native_shifted}, alongside their references.

\begin{deluxetable*}{lcrrrrrlr} 
\tablecaption{Basic information of the entire RRL sample. \label{tab:stars} }
\tablewidth{\textwidth}
\tablehead{
\colhead{GaiaID} & \colhead{Star} & 
\colhead{RA$_{J2000}$} & \colhead{Dec$_{J2000}$} & \colhead{Vmag} & \colhead{Vamp} & \colhead{P} & 
\colhead{Class} & \colhead{Sample}\\
\colhead{(DR2) } &\colhead{ } & 
\colhead{(deg)} &  \colhead{(deg)} & \colhead{(mag)} & \colhead{(mag)} & \colhead{(day)} & 
\colhead{ } & \colhead{ }\\
}
\startdata
4224859720193721856 & AA Aql & 309.5628 &  -2.8903 & 11.831 & 1.275 & 0.3618 & RRab & TW-RRL \\
2608819623000543744 & AA Aqr & 339.0161 & -10.0153 & 12.923 & 1.087 & 0.6089 & RRab & TW-RRL \\
3111925220109675136 & AA CMi & 109.3299 &   1.7278 & 11.558 & 0.965 & 0.4763 & RRab & TW-RRL \\
1234729400256865664 & AE Boo & 221.8968 &  16.8453 & 10.651 & 0.423 & 0.3149 & RRc  & TW-RRL \\
2150632997196029824 & AE Dra & 276.7780 &  55.4925 & 12.474 & 0.799 & 0.6027 & RRab & Lit-RRL \\
\enddata
\tablecomments{Identification, coordinates, average visual magnitude (Vmag), visual amplitude (Vamp), 
period (P), classification, and sample of the RRL stars. 
Table \ref{tab:stars} is published in its entirety in machine-readable format.
A portion is shown here for guidance regarding its form and content.
}
\end{deluxetable*}

\begin{deluxetable*}{lrrrrrrrrr} 
\tablecaption{RRL abundances adopted from the literature. \label{tab:literature_native_shifted} }
\tablewidth{\textwidth}
\tablehead{
\colhead{GaiaID} & 
\colhead{[Fe/H]$_{\text{o}}$} & \colhead{[Mg/Fe]$_{\text{o}}$} & \colhead{[Ca/Fe]$_{\text{o}}$} & \colhead{[Ti/Fe]$_{\text{o}}$} & 
\colhead{[Fe/H]} & \colhead{[Mg/Fe]} & \colhead{[Ca/Fe]} & \colhead{[Ti/Fe]} & \colhead{Reference}\\
\colhead{(DR2) } & 
\colhead{ } &  \colhead{ } & \colhead{ } & \colhead{ } & 
\colhead{ } & \colhead{ } & \colhead{ } & \colhead{ } & \colhead{ }
}
\startdata
15489408711727488  & -2.59  &	      & 0.34 & 0.43 & -2.59 &      & 0.34 & 0.43 & C17 \\
15489408711727488  & -2.48  &	0.48  & 0.45 & 0.43 & -2.68 & 0.61 & 0.44 & 0.38 & C95 \\
15489408711727488  & -2.47  &	      & 0.29 &      & -2.48 &      & 0.39 &      & L96 \\
15489408711727488  & -2.19  &	0.48  & 0.29 & 0.80 & -2.42 & 0.36 & 0.32 & 0.91 & P15 \\
234108363683247616 & -0.28  &	      & 0.14 &      & -0.28 &      & 0.09 &      & F96 \\
\enddata
\tablecomments{Identification, iron and $\alpha$-element abundances in both their original (subscript \emph{o}) scale and in our scale
for the Lit-RRL sample. 
Table \ref{tab:literature_native_shifted} is published in its entirety in machine-readable format.
A portion is shown here for guidance regarding its form and content.
References and number of stars in common with the TW-RRL sample are:
C17, 13: \citet{2017ApJ...835..187C};
C95, 8:  \citet{1995AJ....110.2319C};
F10, 1:  \citet{2010AJ....140.1694F};
F11, 5:  \citet{2011ApJS..197...29F};
F96, 4:  \citet{1996AandA...312..957F},
G14, 2:  \citet{2014ApJ...782...59G};
L96, 8:  \citet{1996ApJS..103..183L};
P15, 8:  \citet{2015MNRAS.447.2404P};
S17, 5:  \citet{2017ApJ...848...68S}.
}
\end{deluxetable*}

\citet{2010AJ....140.1694F} investigated the chemical abundances of metal-poor field red HB stars, in our same
metallicity and $\alpha$-element scale. We found that two of their stars were later classified as RRL. One of them is
already in the TW-Lit sample, and the other was added to the Lit-RRL sample. We adopted the data for the remaining HB 
stars as the Lit-HB sample, with 46 stars. The complete sample of RRL and HB stars in this work will be referred to 
as the RRL+HB sample.


\section{Chemical abundance measurements} \label{ch:measurements}
We have applied the same iron line list and LTE line analysis described in \citet{2021ApJ...908...20C}. In brief, equivalent
widths were measured manually with the {\em splot} IRAF. We only considered lines with equivalent widths 
between 15 and 150 m\AA{} in order to avoid spurious measurements and saturated lines. We derived the effective temperature (T$_{\text{eff}}$), surface gravity (log(g)), microturbulent 
velocity ($\xi_{\text{t}}$), and metallicity ([Fe/H]) for each atmosphere using the equivalent widths of the neutral
and single-ionized iron lines. For this, we followed the method of iteratively changing the atmospheric parameters in order to achieve excitation equilibrium of FeI lines (T$_{\text{eff}}$), ionization equilibrium between FeI and FeII lines (log(g)), and no 
trend between the abundance of each individual FeI line against its respective reduced equivalent width ($\xi_{\text{t}}$). This process
was done using the 2019 release of {\textsc{Moog}}\footnote{The code and documentation can be found at 
\url{https://www.as.utexas.edu/~chris/moog.html}} \citep{1973ApJ...184..839S}, the {\textsc{Moog}} wrapper 
{\textsc{pyMOOGi}}\footnote{The code and documentation can be found at \url{https://github.com/madamow/pymoogi}} 
developed by M. Adamow, and an interpolated grid of $\alpha$-enhanced  ([$\alpha$/Fe] = 0.4 dex){\textsc{ATLAS9}} model atmospheres 
\citep{2003IAUS..210P.A20C}. The adopted atmospheric values for each individual measurement are shown in 
Tab.~\ref{tab:individual_atmospheres}.

Once the final model atmosphere was constrained, the abundances of the $\alpha$-elements were computed from the 
equivalent widths of their lines. The line list is shown in Tab.~\ref{tab:alpha_linelist}, alongside the reference
for their excitation potential (EP) and oscillator strength (log(gf)). As with iron, only lines with equivalent widths between 
15 and 150 m\AA{} were considered. Solar abundance values were adopted 
from \citet{2009ARAandA..47..481A}. In the case where more than one line was available for a given chemical species, 
the median\footnote{In this work, we employ the $\mu$ and $\sigma$ characters to denote, respectively, the median and the median 
absolute deviation.} value was adopted. 

A single RRL can undergo changes as large as 1000 K in effective temperature and 1 dex in log(g) \citep{2011ApJS..197...29F}. 
Thus, the robustness of a given method of abundance determination can be assessed by its capacity to recover coherent 
values across the pulsation phase. Similarly, the difference between repeated measurements is a reliable determination of
the uncertainty of the measurements. With this in mind, we computed the uncertainties for iron and individual $\alpha$-elements by 
taking the median absolute deviation between multiple
measurements for the same star both in the TW-RRL and Lit-RRL samples. For the TW-RRL sample, this allowed us to determine the typical uncertainty
of each chemical species in each spectrograph, which we adopted for the stars with a single measurement. For stars with a single
measurement in the Lit-RRL sample, we adopted a fixed uncertainty of 0.10 dex for iron, and 0.15 for each $\alpha$-element.
We averaged the abundances of TiI and TiII in order to derive a total [Ti/Fe] ratio. The [TiI/Fe] and [TiII/Fe] are on average 
shifted by 0.05 dex in our data. Any disagreements between TiI and TiII abundances are reflected in the uncertainties for each star. 

To compute the total [$\alpha$/H] abundance, we took the median of [Mg/H], [Ca/H], [TiI/H], and [TiII/H]
according to their availability and weighted by their uncertainties. The median absolute deviation between the different 
$\alpha$-elements was adopted as the uncertainty in the total [$\alpha$/H] abundance. Finally, we subtracted the iron abundance
from [$\alpha$/H] for each individual star to arrive at the final [$\alpha$/Fe] value.

\subsection{Verification of the T$_{\text{eff}}$ scale}
The atmospheric parameter that most strongly affects the determination of chemical abundances is the effective temperature. As described above,
it is essential that the methodology employed for chemical abundance analysis be capable of recovering coherent abundances across the pulsation 
cycle, i.e. at different values of T$_{\text{eff}}$. It is already known that spectroscopic studies of RRL in both low and high resolution 
can achieve excellent precision at random phases \citep[e.g.][]{2011ApJS..197...29F,2021ApJ...908...20C}. 

In order to have a sanity check indepedent of spectroscopy, we applied the photometric T$_{\text{eff}}$ calibration of 
\citet{1999A&AS..140..261A,2001A&A...376.1039A} to a subsample of RRL with V- and K-band photometry. The V-K color was chosen because it is the 
least affected by uncertainties in the temperature and provides very stable results \citep{2000ASPC..203..176C,2003LNP...635...85B}. Photometric
T$_{\text{eff}}$ relations have a limited applicability to the RRLs because these stars cover a wide range of metallicities and moderately high temperatures
\citep[see e.g. Table 1 in][]{1999A&AS..140..261A}. As the RRL are variable stars with continuously changing colors, the application of these calibrations requires either simultaneous or well sampled light curves in both optical and near-infrared bands. Moreover, the phasing itself requires very good determinations of both period and reference epoch. An added difficulty is that, in order to adopt photometric temperatures in a chemical abundance analysis, both photometric and spectroscopic data must be acquired for the same phase. Unfortunately, obtaining all the necessary data for these paired observations is not trivial. 

\begin{figure}
\includegraphics[width=\columnwidth]{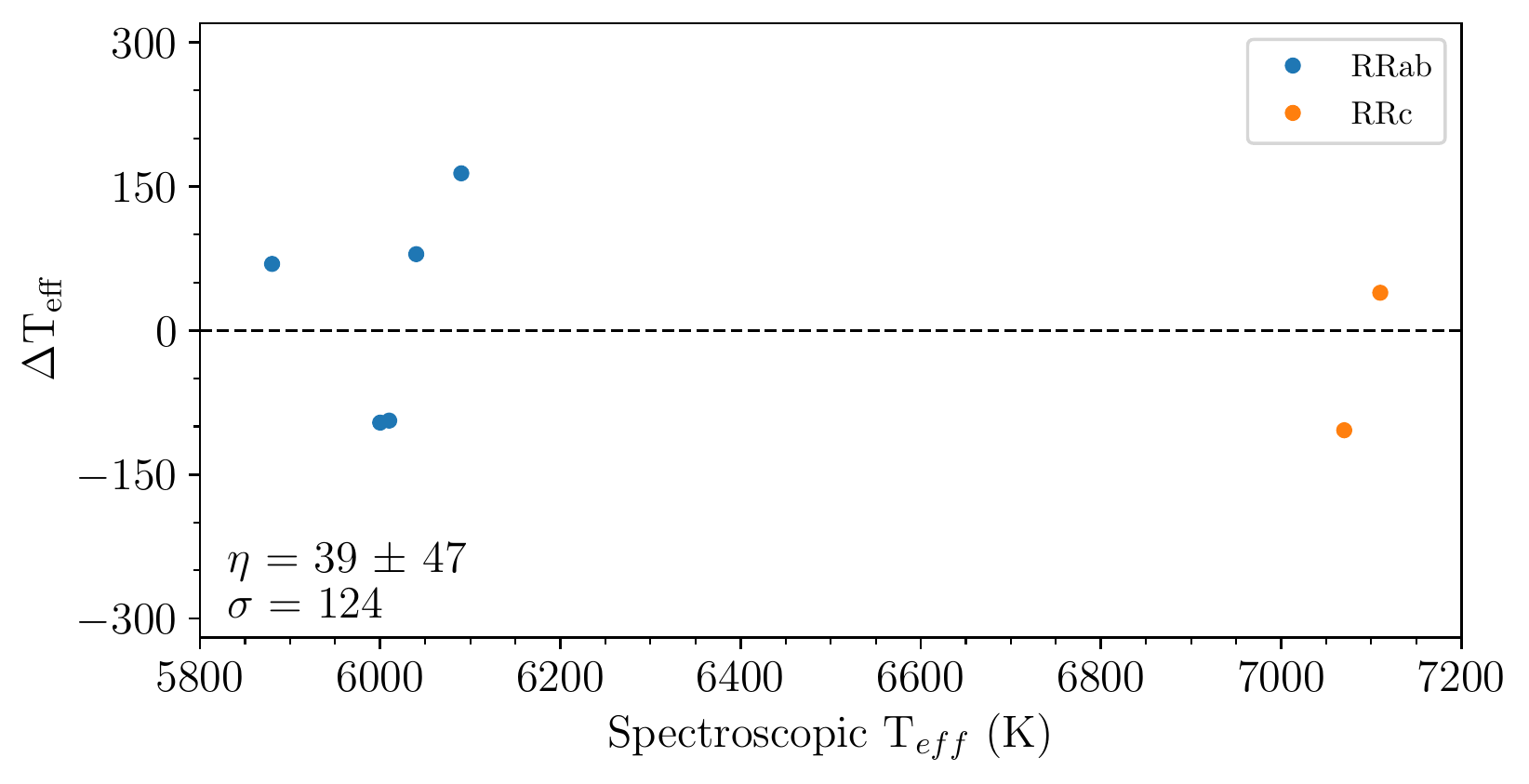}
\caption{Difference between the photometric and spectroscopic effective temperatures $\Delta$T$_{\text{eff}}$ for seven measurements
of three RRLs. The median ($\eta$) and median absolute deviation ($\sigma$) of the difference is shown on the lower left. See text for details.
\label{fig:Teff_compare}}
\end{figure}

Fortunately enough, we found well sampled V-K colors curves for three variables with a total of seven spectra in the TW-RRL sample: DH Peg (RRc, [Fe/H] = -1.36, two spectra), 
VY Ser (RRab, [Fe/H] = -1.96, three spectra), and W Tuc (RRab, [Fe/H] = -1.90, two spectra). The photometry was taken from 
\citet{1988ApJ...326..312J,1988ApJS...67..403B,1989ApJS...69..593L,1990A&AS...85..865C,1990MNRAS.247..287F,1992ApJ...396..219C}. We found a very
good agreement between photometric and spectroscopic estimates, with residuals displaying a median $\eta$=39$\pm$47 K, and median absolute deviation $\sigma$=124 K 
(Fig.~\ref{fig:Teff_compare}).

\begin{deluxetable*}{lcrrrrrrrr} 
\tablecaption{Atmospheric parameters derived in this work for each individual measurement. \label{tab:individual_atmospheres}}
\tablewidth{\textwidth}
\tablehead{
\colhead{GaiaID} & \colhead{Spectrograph} &
\colhead{T$_{\text{eff}}$} & \colhead{log(g)} &  \colhead{$\xi_{\text{t}}$} & 
\colhead{[FeI/H]}  & \colhead{N$_{\text{FeI}}$} & \colhead{[FeII/H]} & \colhead{N$_{\text{FeII}}$}  & \colhead{N$_{\text{stack}}$} \\
\colhead{(DR2)} & \colhead{ } & 
\colhead{(K)} & \colhead{(dex)} & \colhead{(km s$^{-1}$)} & 
\colhead{(dex)} & \colhead{ } & \colhead{(dex)} & \colhead{ } & \colhead{ } 
}
\startdata
4224859720193721856 & SALT   & 6610$\pm$130 & 2.70$\pm$0.12 & 2.52$\pm$0.08 & -0.34$\pm$0.24 & 206 & -0.34$\pm$0.22 & 39 & 1 \\
4224859720193721856 & Subaru & 6470$\pm$110 & 2.62$\pm$0.10 & 2.42$\pm$0.08 & -0.49$\pm$0.17 & 137 & -0.49$\pm$0.16 & 22 & 1 \\
2608819623000543744 & UVES   & 5840$\pm$160 & 1.52$\pm$0.06 & 3.51$\pm$0.25 & -2.31$\pm$0.10 &  37 & -2.31$\pm$0.12 & 13 & 1 \\
3111925220109675136 & SALT   & 7090$\pm$180 & 3.01$\pm$0.15 & 3.04$\pm$0.16 &  0.24$\pm$0.24 & 146 &  0.24$\pm$0.21 & 19 & 1 \\
1234729400256865664 & HARPS  & 6630$\pm$150 & 2.04$\pm$0.08 & 2.79$\pm$0.09 & -1.62$\pm$0.14 &  64 & -1.62$\pm$0.10 & 25 & 2 \\
\enddata
\tablecomments{Atmospheric parameters for each indvidual measurement of the TW-RRL sample. The columns N$_{\text{FeI}}$ and N$_{\text{FeII}}$
contain the number of adopted FeI and FeII lines, respectively. Column N$_{\text{stack}}$ shows the number of individual exposures that
are stacked in order to obtain the measurement. See text for details.
Table \ref{tab:individual_atmospheres} is published in its entirety in machine-readable format.
A portion is shown here for guidance regarding its form and content.}
\end{deluxetable*}

\begin{deluxetable}{lllrr} 
\tablecaption{List of $\alpha$-element transitions adopted in this work.
\label{tab:alpha_linelist}}
\tablewidth{\columnwidth}
\tablehead{
\colhead{Wavelength} & \colhead{Species} & \colhead{EP} & \colhead{log(gf)} & \colhead{Reference} \\
\colhead{(\AA{}{}) } & \colhead{ } & \colhead{(eV)} & \colhead{(dex)} & \colhead{ } \\
}
\startdata
3829.36  &  12.0  &  2.709  &  -0.227  &  NIST \\
4571.10  &  12.0  &  0.000  &  -5.620  &  NIST \\
4702.99  &  12.0  &  4.346  &  -0.440  &  NIST \\
5172.68  &  12.0  &  2.712  &  -0.393  &  NIST \\
5183.60  &  12.0  &  2.717  &  -0.167  &  NIST \\
\enddata
\tablecomments{
Table \ref{tab:alpha_linelist} is published in its entirety in machine-readable format.
A portion is shown here for guidance regarding its form and content.
References -- 
NIST: \url{https://www.nist.gov/}, 
LAW2013:  \citet{2013ApJS..205...11L},
WOO2013:  \citet{2013ApJS..208...27W}.
}
\end{deluxetable}

\subsection{Validation of the $\alpha$-element abundance scale}
\begin{figure}
\includegraphics[width=\columnwidth]{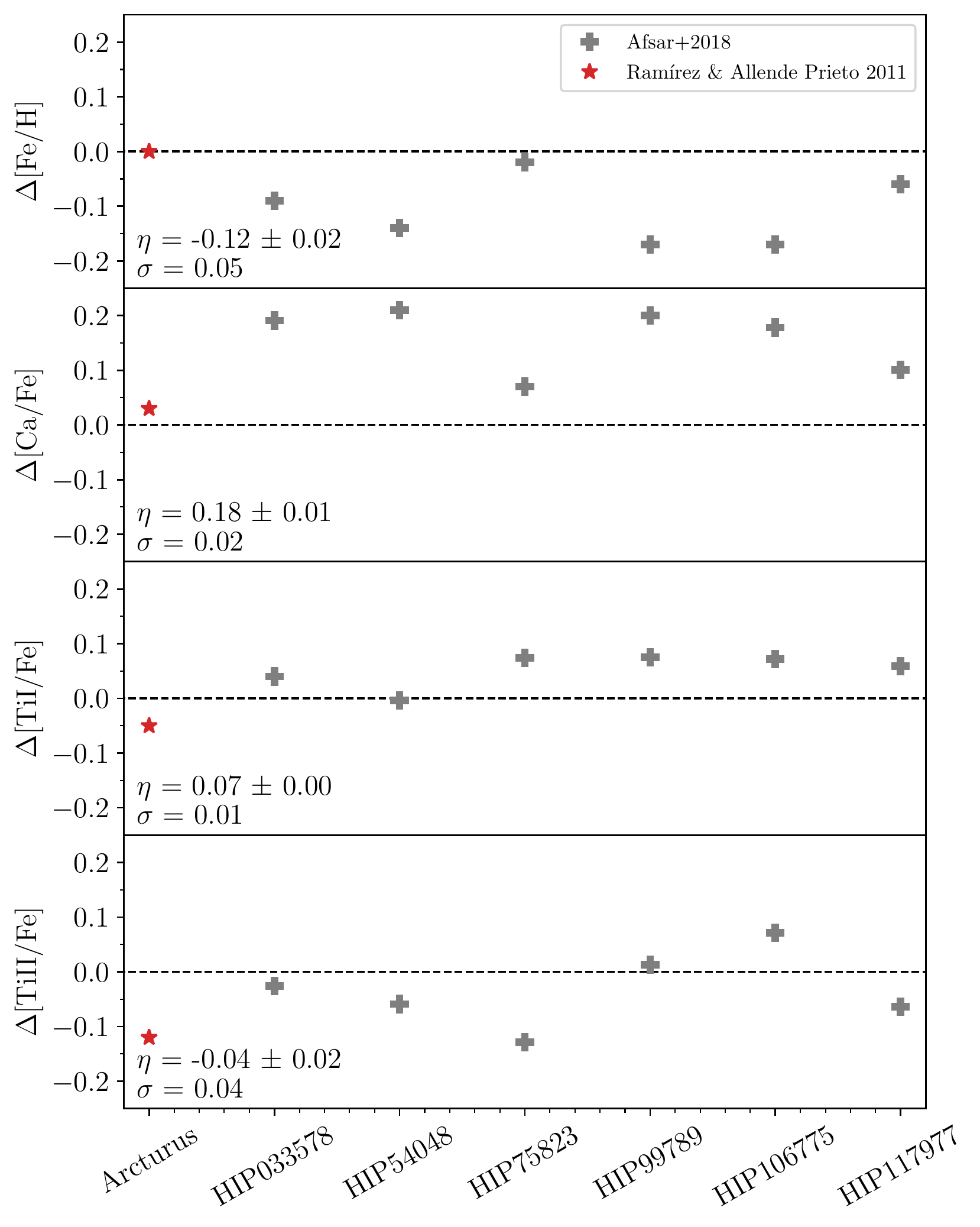}
\caption{Difference between the iron and $\alpha$-element abundances derived in 
this work and those of \citet{2018AJ....155..240A} for a sample of six HB stars,
 shown as grey crosses. The median ($\eta$) and median absolute deviation ($\sigma$) 
of the difference is shown in each panel. A red star shows the same difference for Arcturus 
\citep{2011ApJ...743..135R}.
\label{fig:scale_comparison}}
\end{figure}

The validation of our metallicity scale was performed in \citet{2021ApJ...908...20C}. For the validation of our
 $\alpha$-element abundance scale, we performed three tests. First, we analysed one high SNR ($\approx$ 350), 
high dispersion (R=115000) spectrum for Arcturus collected with HARPS. We found the atmospheric parameters 
T$_{\text{eff}}$ = 4350$\pm$60 K, log(g) = 1.65$\pm$0.07, $\xi_{\text{t}}$ = 1.75$\pm$0.04 kms$^{-1}$, and chemical 
abundances\footnote{The listed uncertainties in the chemical abundances for Arcturus are only due to the
uncertainties in T$_{\text{eff}}$, log(g), and $\xi_{\text{t}}$.} [FeI/H] = -0.52$\pm$0.06, [FeII/H] = -0.52$\pm$0.20, 
[Ca/Fe] = 0.08$\pm$0.14, [TiI/Fe] = 0.32$\pm$0.20, and [TiII/Fe] = 0.33$\pm$0.09 dex. These results are in excellent agreement with 
\citet{2011ApJ...743..135R}. Indeed, the difference in the abundances 
is 0.00, 0.03, -0.05, -0.12 for Fe, Ca, TiI, and TiII, respectively. Unfortunately, the Mg lines in
our line list, optimized for hotter stars, were all saturated in the much colder atmosphere of Arcturus. Second, we 
performed the same analysis on
six red HB stars investigated by \citet{2018AJ....155..240A}. These stars are only slightly colder than the RRL. The results are shown in 
Fig.~\ref{fig:scale_comparison}, and once again they agree quite well with literature estimates. Third, we made a 
comparison using directly the equivalent widths of iron and $\alpha$-element for several pairs of stars at similar 
effective temperature. The analysis of these paired spectra is discussed in Appendix \ref{ap:EW_comparisons}.

We verified that NLTE corrections do not change the conclusions of our investigation (Sect.~\ref{ch:results}).
As most works in the literature do not make use of such corrections, we opted to not apply them in order to better
compare our results to previous ones. Ca, TiI, and TiII display a few lines that appear in all metallicity regimes 
and allowed us to verify that the averages for each species are not affected by systematics between lines. For
Mg, no individual line is measurable in the entire metallicity range, but different lines have superposed metallicity 
regimes, e.g. one line appears in stars from metal-poor to -intermediate, and another from metal-intermediate 
to -rich. These considered together exhibit a coherent trend with each other, and with both Ca and Ti. We refer 
the reader to Appendix \ref{ap:NLTE_individuallines} for a detailed discussion of both the NLTE corrections and the 
behavior of individual lines.

\section{Results and discussion} \label{ch:results}
\subsection{The individual species} \label{ch:results_individual}
\begin{figure}
\includegraphics[width=\columnwidth]{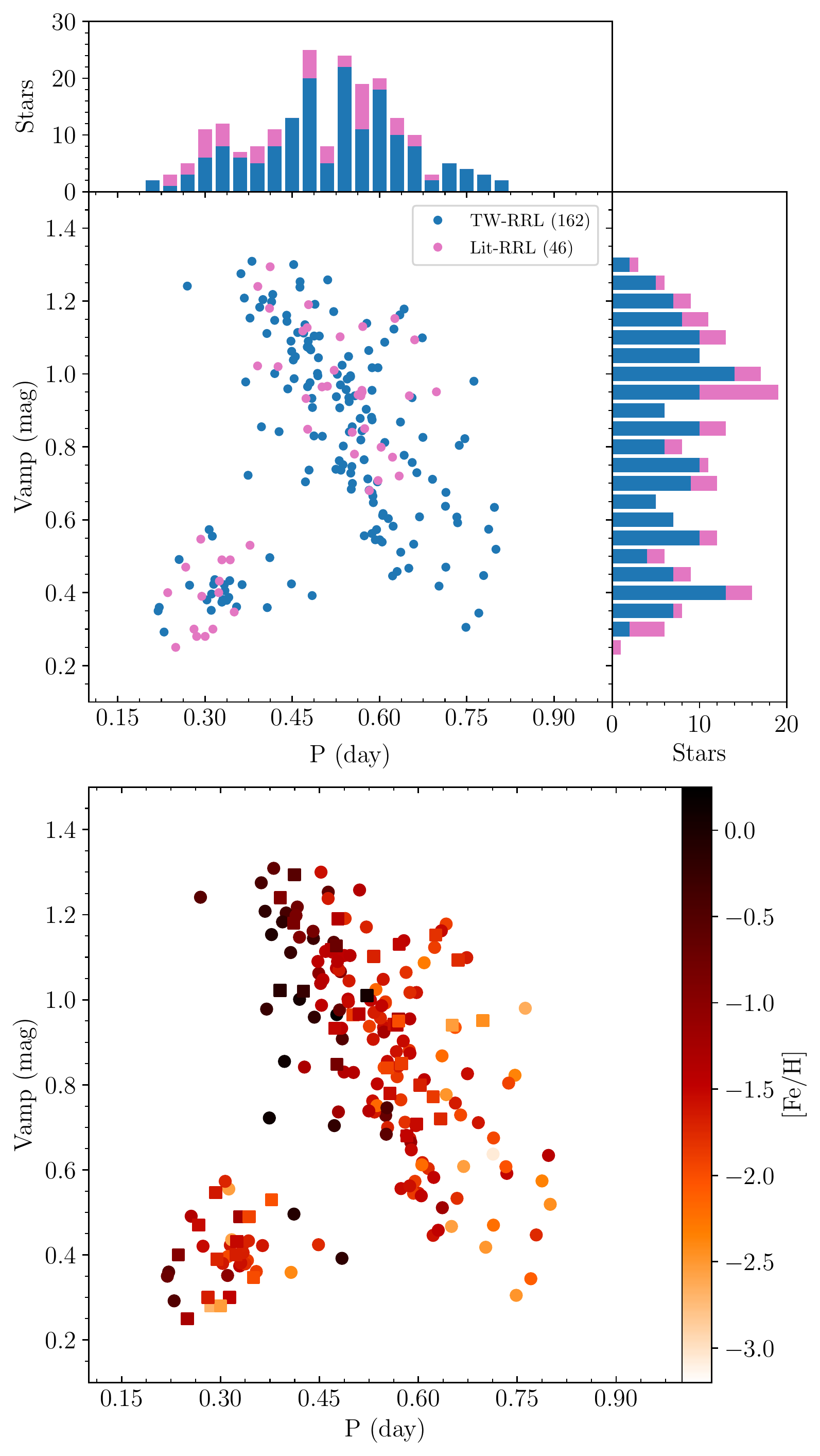}
\caption{\emph{Top:} Bailey diagram for the TW-RRL sample (blue), and the Lit-RRL sample
(pink). The top and left histograms show the distribution of period and V-band amplitudes, 
respectively. \emph{Bottom:} Bailey diagram colored by metallicity according to the color
bar on the right. The TW-RRL and the Lit-RRL stars are marked with circles and squares,
respectively.
\label{fig:sample_details}}
\end{figure}

The final metallicity, individual $\alpha$-element abundance, and total [$\alpha$/Fe] 
abundance for each RRL and HB star is shown in Tab.~\ref{tab:results_abundances}. 
The coverage in pulsational amplitude and period of the full RRL sample is shown in the 
top panels of Fig.~\ref{fig:sample_details}. The bottom panel of the same figure shows 
the same sample, but color-coded according to metallicity (see the color bar on the 
right side).

Fundamental RRLs located in the high-amplitude, short-period (P $\lesssim$ 0.48 day) region, 
i.e. the so-called HASP region, are confirmed to be more metal-rich than -1.5 dex, as 
expected from low resolution spectra and globular cluster metallicities 
\citep[see ][]{2015ApJ...798L..12F,2017A&A...599A.125F}. The precision of the current iron 
abundances strenghtens the evidence that the HASPs trace quite well the transition from 
fundamental to first overtone RRLs. The RRc seems to show a similar trend: their metal-rich 
tail is traced by short period variables (P$\le$0.27 day), although their luminosity amplitudes 
have typical RRc values. However, the number of metal-rich RRc variables is still too limited 
to constrain their pulsation properties close to the blue (hot) edge of the RRL instability strip.

\begin{figure}
\includegraphics[width=\columnwidth]{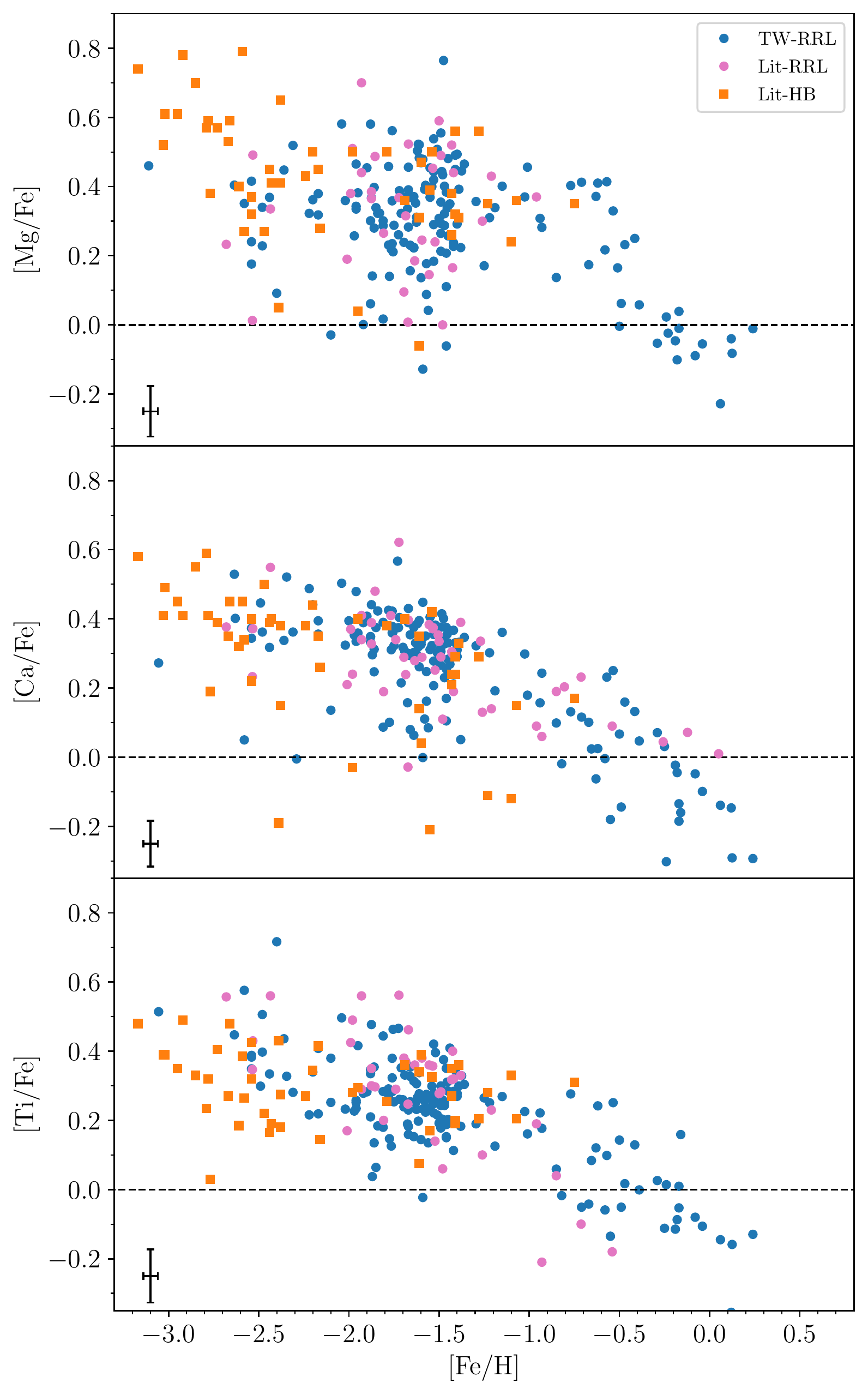}
\caption{Abundances of Mg, Ca, and Ti for the TW-RRL (blue dots), Lit-RRL (pink
dots), and Lit-HB (orange squares) samples. The error bars in the lower left 
corner of each panel display typical uncertainties.
\label{fig:alpha_comparison_RRL}}
\end{figure}

The [X/Fe] versus [Fe/H] plane for each $\alpha$-element of interest is shown in
Fig.~\ref{fig:alpha_comparison_RRL} with both the RRL and the Lit-HB samples. 
Several interesting features are visible and worth being discussed in 
detail. 

\begin{deluxetable*}{lrrrrrr}
\tablecaption{Abundances for the RRL and Lit-HB samples.
\label{tab:results_abundances}}
\tablewidth{\columnwidth}
\tablehead{
\colhead{Gaia ID} & \colhead{[Fe/H]}  & \colhead{[Mg/Fe]} & \colhead{[Ca/Fe]} &
                    \colhead{[Ti/Fe]} & \colhead{[$\alpha$/Fe]} & \colhead{Sample} \\
\colhead{(DR2) }  & \colhead{(dex)}   & \colhead{(dex)}   & \colhead{(dex)}   & 
                    \colhead{(dex)}   & \colhead{(dex)}   & \colhead{ }
}
\startdata
15489408711727488  & -2.53$\pm$0.09 & 0.49$\pm$0.20 & 0.37$\pm$0.20 & 0.43$\pm$0.20 & 0.37$\pm$0.09 & Lit-RRL \\
53848448829915776  & -1.41$\pm$0.03 & 0.56$\pm$0.15 & 0.29$\pm$0.15 & 0.20$\pm$0.15 & 0.26$\pm$0.06 & Lit-HB  \\
77849374617106176  & -1.78$\pm$0.02 & 0.14$\pm$0.03 & 0.10$\pm$0.04 & 0.15$\pm$0.08 & 0.13$\pm$0.02 &  TW-RRL \\
80556926295542528  & -1.88$\pm$0.09 & 0.58$\pm$0.12 & 0.35$\pm$0.11 & 0.35$\pm$0.10 & 0.43$\pm$0.09 &  TW-RRL \\
234108363683247616 & -0.26$\pm$0.02 &               & 0.04$\pm$0.20 &               & 0.04$\pm$0.06 & Lit-RRL \\
\enddata
\tablecomments{Table \ref{tab:results_abundances} is published in its entirety in machine-readable format.
A portion is shown here for guidance regarding its form and content.
}
\end{deluxetable*}

i) {\em Similar trends for RRL and HB stars} -- The targets plotted in Fig.~\ref{fig:alpha_comparison_RRL} come from the same evolutionary path. The current 
empirical and theoretical evidence indicates that blue HB, 
RRL and red HB stars are old (t$\ge$ 10 Gyr), low-mass 
(M$\approx$0.50-0.95M$_\odot$) stars in their central helium burning phase. They have very similar 
helium core masses ($\sim$0.50M$_\odot$) and their key difference is in the envelope mass. 
A steady decrease in the envelope mass causes a systematic increase in the effective temperature
moving the stellar structure from the red HB to the blue HB, passing through the RRL instability strip.        
There is evidence that some RRLs are the aftermath of close binary evolution and could be younger
objects \citep{2012Natur.484...75P,2017MNRAS.466.2842K}. However, the fraction of RRLs in 
binary systems is of the order of a few percent \citep{2015MNRAS.449L.113H,2019A&A...623A.116K,2019MNRAS.487L...1P}, and 
the figure for systems with mass transfer is likely to be even more modest. In all our spectroscopic 
investigations with the present sample, we found no evidence of binarity.

ii) {\em Similar slopes for Mg, Ca, and Ti} -- The investigated species display 
a well defined slope when moving from the metal-poor to the metal-rich regime. 
The steady decrease in $\alpha$-enhancement is more clear in Ti and Ca for which the variation is 
of the order of $\sim$0.6 dex, but it is also present in Mg. Very metal-poor 
({Fe/H]$\le$-2.2) RRLs are strongly enhanced in $\alpha$-elements 
([$\alpha$/Fe]$\sim$0.4 to 0.5), while those approaching solar iron abundance are 
depleted in $\alpha$-elements ([$\alpha$/Fe]$\sim$-0.2 to -0.3, see also \citealt{2020MNRAS.492.3408P}).     

iii) {\em Similar dispersion for Ca and Ti} -- Both Ca and Ti display trends in tight agreement and
can be considered the same within uncertainties. Their scatter remains of the order of 0.4 dex over
the entire metallicity range and appears to be intrinsic, because it is over 3 times larger than the typical 
 errors (see typical error bars in the bottom left corner of Fig.~\ref{fig:alpha_comparison_RRL}). 

The value of the plateau for these two species is not significantly different in our data. It bears
mentioning that a disagreement between Ti and Ca in a given investigation may be a consequence of the adopted
atomic lines and their transition parameters. Indeed, updated transition parameters derived from
laboratory studies are only available for titanium. For calcium, a variety of parameters can be found
ranging from laboratory studies dating back to half a century ago to astrophysical determinations that use the 
Sun or nearby bright stars as a reference. Differences in the adopted lines and their oscillator strengths can result in
abundances with variations of the order of 0.2 dex \citep{2010A&A...511A..56P}. This limitation 
coupled with smaller sample sizes may have created difficulties in detecting the $\alpha$-element 
depletion we observe in metal-rich RRLs \citep[e.g.][]{2013RAA....13.1307L}.

iv) {\em Larger dispersion for Mg} -- Mg shows a large dispersion at metallicities lower than 
[Fe/H]$\lesssim$-1.2. Measurement difficulties play a role in the scatter.
The number of Mg lines is very limited, with mostly strong lines that easily saturate and must
be discarded. This means that for several stars the Mg abundance is computed using only
one or two transitions. Meanwhile, our spectra typically contained five to 15 lines of varied strengths
for Ca and for Ti. As with calcium, magnesium lacks updated transition parameters from laboratory studies.
An intrinsic spread, independent of the number of lines and their quality, but rather due to the mechanisms of Mg nucleosynthesis may be present and it is discussed in
Sect.~\ref{ch:results_galactic_chemistry}.

\subsection{Comparison with different Galactic components}

The range in metallicity covered by field RRLs is significantly larger than any 
other similar datasets in the literature. This is strikingly clear in a comparison with 
typical stars of different Galactic components, as shown in Fig.~\ref{fig:alpha_comparison_bulgehalodisks}. The RRLs cover the metal-poor
([Fe/H]$\leq$-2.5) region of the Halo and Bulge, but they also extended to super solar
metallicities like the dwarfs and giants in the Thin and Thick Disks.

\begin{figure*}
\includegraphics[width=2\columnwidth]{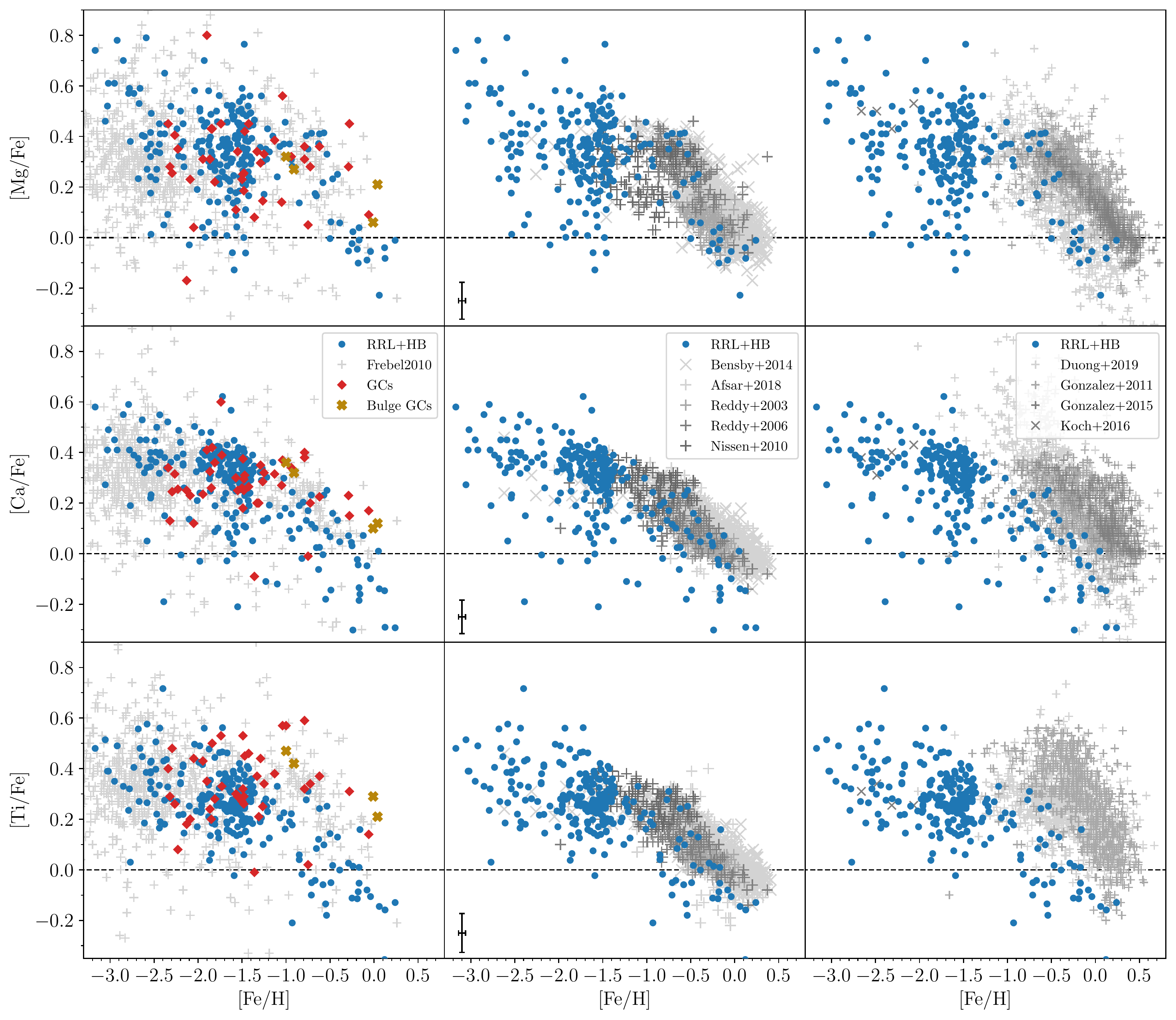}
\caption{Abundances of Mg, Ca, and Ti for the RRL+HB sample, compared to three typical
Galactic populations.
\emph{Left:} Globular clusters and Halo field stars. \emph{Middle:} Thin and thick disk field 
stars. \emph{Right:} Bulge field stars. 
References -- 
GCs: \citet{2005AJ....130.2140P,2009A&A...505..139C,2009A&A...505..117C,2010ApJ...712L..21C},
Bulge GCs: \citet{2011A&A...530A..54G},
The remaining references are labeled according to the first author and year of publication:
\citet{2010AN....331..474F},
\citet{2014A&A...562A..71B},
\citet{2003MNRAS.340..304R},
\citet{2006MNRAS.367.1329R},
\citet{2018AJ....155..240A},
\citet{2010A&A...511L..10N},
\citet{2019MNRAS.486.3586D},
\citet{2011A&A...530A..54G},
\citet{2015A&A...584A..46G},
\citet{2016A&A...587A.124K}.
\label{fig:alpha_comparison_bulgehalodisks}}
\end{figure*}

As suggested by \citet{2010A&A...511L..10N} and more recently by the near-infrared APOGEE survey
\cite{2018ApJ...852...49H}, there is evidence of a bimodal distribution in the [Mg/Fe] versus [Fe/H] plane
for metallicities between -1.50 and -0.50 dex. 
This bimodality is not observed in our data, suggesting that it may an age-related phenomenon. 
In this same Mg vs iron plane, field RRLs, field stars and GCs attain quite similar values in 
the Halo, with a dispersion that allows only modest claims about a slope. On the other hand, 
field RRLs display a well defined slope in Ca and in Ti when moving from [Fe/H]$\approx$-3.2 to -1.3,
while field stars and GCs display an almost constant value at $\approx$0.3 dex. 

\subsection{Comparison with nearby dwarf galaxies}
The current Cold Dark Matter cosmological simulations suggest that the
Halo formed from the aggregation of protogalactic fragments -- small
galaxies form first and then merge to form larger galaxies 
\citep{1986ApJ...303...39D,2005ApJ...635..931B,2019MNRAS.485.2589M}. 
The discovery of stellar streams and the merging of a massive
dwarf galaxy like Sagittarius \citep{1994Natur.370..194I}, Gaia Enceladus \citep{2018Natur.563...85H}
and Sequoia \citep{2019MNRAS.488.1235M} provided further support to this
hierarchical mechanism. Metallicity distribution functions can provide
solid quantitative constraints on the mass assembly of the Galactic Halo
\citep{2017A&A...599A.125F}, therefore, we also compared the current
$\alpha$-element abundances with similar abundances for RGs in nearby dwarf
galaxies.

The data plotted in Fig.~\ref{fig:alpha_comparison_dgals} display the same comparison of 
Fig.~\ref{fig:alpha_comparison_bulgehalodisks}, but for $\alpha$-element abundances of 
individual RG stars in both classical dwarf galaxies and Ultra Faint Dwarfs (UFDs). Note that 
in this comparison we are only taking into account measurements based on HR spectra.

The samples for Sagittarius and Fornax are marked by red stars and goldenrod diamonds,
respectively. The comparison indicates a remarkable agreement in the metal-rich tail 
([Fe/H]$\ge$-1.2) between Halo RRLs and RGs in Sagittarius and, in particular, the $\alpha$-poor
RRLs approaching solar iron abundance. The trend in the three different $\alpha$-elements are 
similar across a range in iron abundance of over 1 dex, with marginal variations. In the case 
of the RGs in Fornax, however, the agreement with the RRL+HB sample is only present for metallicities 
between -1.3 to -1.8 dex. Indeed, the bulk of Fornax RGs are on overage more $\alpha$-poor than our 
sample.

A good agreement can also be seen in the metal-poor regime ([Fe/H]$\leq$-2.2) between the RRL+HB 
sample and RGs in UFDs \citep[grey crosses, data from][]{2013ApJ...767..134V}. This is quite interesting, 
because the age distribution in UFDs is narrower when compared to classical dSph galaxies. However, 
recent spectroscpic measurements are suggesting that the chemical abundance distributions of RGs in UFDs
is inhomogeneous \citep{2008ApJ...688L..13K,2014ApJ...789..148W}. The empirical framework becomes even more complex for the more massive
dwarf galaxies because they exhibit a broad range of star formation histories and chemical enrichment histories 
\citep{2009ARA&A..47..371T}. Indeed, the $\alpha$-element abundances for classical dwarf galaxies (light blue
squares) have a dispersion in $\alpha$ abundances, at fixed iron content, that is significantly larger than for 
our sample. Note that this trend is also caused by the fact that the current $\alpha$-element abundances 
are restricted, due to an observational bias, to bright RGs in nearby dwarf galaxies. This means that they typically cover 
a broad range in age. \citet{2015A&A...580A..18F} has recently addressed this issue and found that old 
and intermediate-age stellar populations in the Carina dSph galaxy display a difference of $\sim$0.6 dex
in $\alpha$-element abundances \citep[see also ][]{2008AJ....135.1580K}. Unfortunately, we still lack accurate abundance estimates of 
$\alpha$-elements in truly old stellar tracers (RRL and non-variable HB stars) belonging to 
nearby dwarf galaxies to be compared to our sample.

\begin{figure}
\includegraphics[width=\columnwidth]{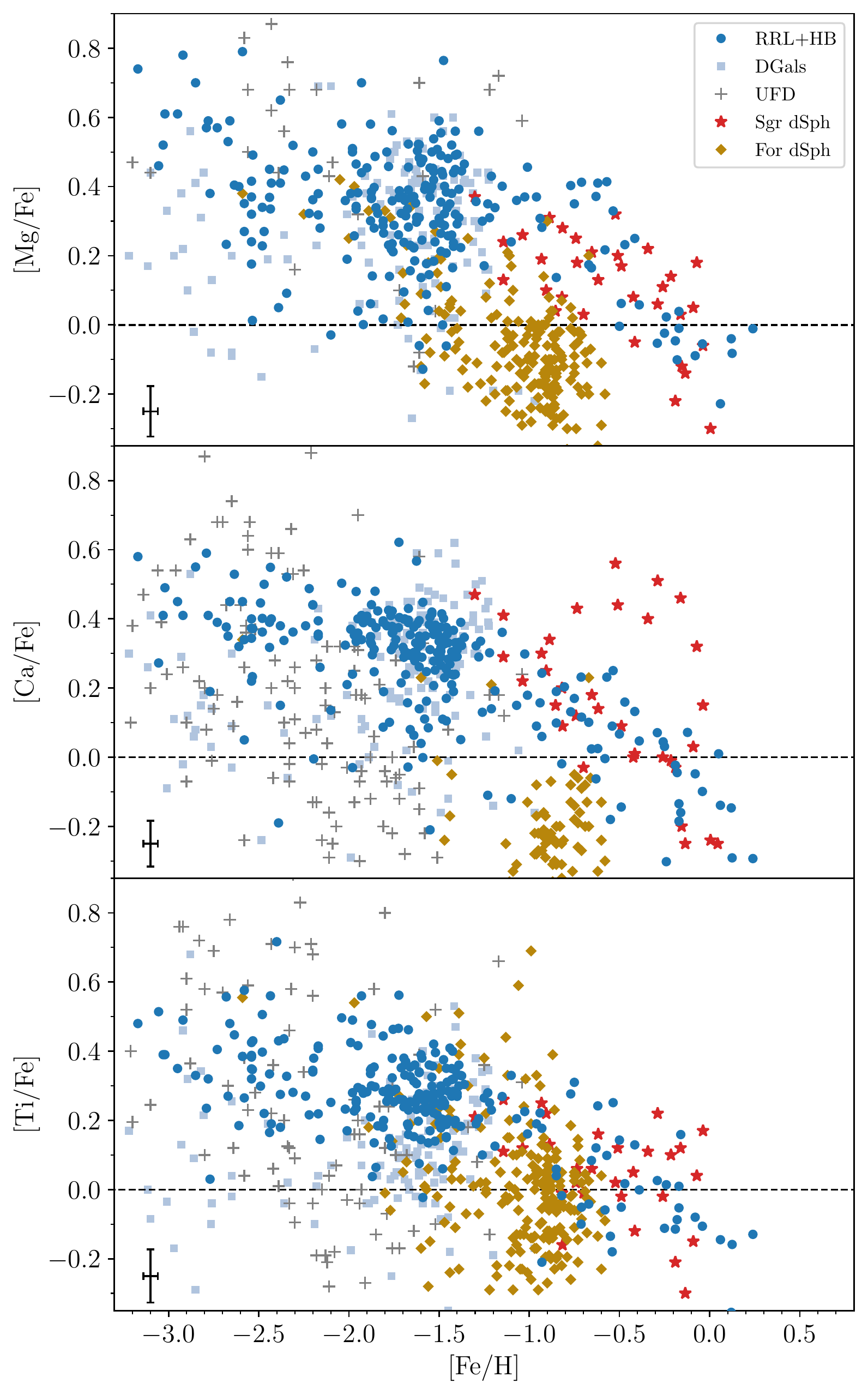}
\caption{Abundances of Mg, Ca, and Ti for the RRL+HB sample, compared to stars in nearby dwarf
galaxies. Red stars: Sagittarius dSph. Goldenrod diamonds: Fornax dSph. Light blue squares: Carina, 
Draco, Leo I, Sculptor, Sextans, and Ursa Minor. Grey crosses: ultrafaint dwarf galaxies.
References -- 
\citet{2001ApJ...548..592S},
\citet{2003AJ....125..707T},
\citet{2005AJ....129.1428G},
\citet{2007A&A...465..815S},
\citet{2009A&A...502..569A}
\citet{2009ApJ...701.1053C},
\citet{2010ApJ...708..560F},
\citet{2010A&A...520A..95C},
\citet{2010A&A...523A..17L},
\citet{2010A&A...524A..58T},
\citet{2013A&A...549A..88S},
\citet{2013ApJ...767..134V},
\citet{2014A&A...572A..82H}.
\label{fig:alpha_comparison_dgals}}
\end{figure}

\subsection{Preliminary circumstantial evidence concerning RRL chemical enrichment} \label{ch:results_galactic_chemistry}
As mentioned in Sect.~\ref{ch:results_individual}, the plateau in the [X/Fe] versus
[Fe/H] plane depends on the quality of the adopted atomic transition parameters. This is the
reason why we are mainly interested in the trends among the different 
$\alpha$-elements. Yet, our results show very good agreement among Mg, Ca, and Ti, including the value
of the plateau, in accordance with both the data from \citet{2010AN....331..474F} shown in the rightmost panels of 
Fig.~\ref{fig:alpha_comparison_bulgehalodisks}. The same agreement is found with the very metal-poor Halo dwarfs and giants investigated by 
\citet{2004A&A...416.1117C}. The logarithmic fits of the [X/Fe] versus [Fe/H] planes agree within errors 
for the three species, and indeed they are nearly identical for Ca and Ti across the whole metallicity 
range (Fig.~\ref{fig:alpha_weighted_averages}, bottom panel).

Magnesium displays a larger spread than Ca and Ti for both the RRL and the Lit-HB samples, and for the
typical populations of each Galactic component (Fig.~\ref{fig:alpha_comparison_bulgehalodisks}). An 
intrinsic spread in Mg and deviations from the trends set by Ca and Ti may be present due to the 
dependence on progenitor mass and metallicity of the Mg yields. In theoretical models, the yields of Ca 
and Ti remain similar for wide range of progenitor masses, but the same cannot be said for Mg. The 
production of the latter significantly increases in progenitors with large stellar masses 
\citep[35 M$_\odot$, see Figure 6 of][]{1997ARA&A..35..503M}. Moreover, the production of Mg, at
fixed stellar mass, depends on the metallicity. Indeed, it shows a marked decrease when moving from 
metal-poor/metal-intermediate to metal-rich progenitors 
\citep[see Figures 2 and 4 in ][]{2006ApJ...653.1145K}. 

However, the current data indicate that [Mg/H], [Ca/H], and [Ti/H] vary in lockstep with one another,
with negligible differences in their dispersion. Taken at face value, this result points to an early chemical enrichment 
that appears to be quite homogeneous for these three species over a wide range in
iron abundance. 

The average [$\alpha$/Fe] versus iron plane is quite homogeneous and properly fit 
by the logarithmic function

\begin{equation}
\text{ [$\alpha$/Fe] = $a$ + $b$ log($c$ - [Fe/H]) }
\label{eq:logfit}
\end{equation}

\noindent with parameters a, b, c, and RMS error as listed in Tab.~\ref{tab:fit_parameters}. The
fit is shown in both panels of Fig.~\ref{fig:alpha_weighted_averages}. We found no
trends in the residuals of this fit against the pulsational properties, i.e.
period and amplitude, of the RRL sample, nor any peculiar behavior when separating 
RRab and RRc stars. Furthermore, we did not observe a significant change in either the 
logarithmic fit parameters nor the residuals when removing Mg. As mentioned in 
Sect.\ref{ch:results_individual}, our spectra have only a small number of Mg lines, 
and so the weighted average of the three species favours Ca and Ti with their nearly 
identical trends. As the $\alpha$-elements considered in this work have different 
formation channels, we included the parameters for the same logarithmic function 
considering each individual chemical species individually in Tab.~\ref{tab:fit_parameters}. 
The corresponding fits are shown in the bottom panel of Fig.~\ref{fig:alpha_weighted_averages}.

\begin{deluxetable}{lllrr} 
\tablecaption{Parameters of the logarithmic fits (Eq.~\ref{eq:logfit}).
\label{tab:fit_parameters}}
\tablewidth{\columnwidth}
\tablehead{
\colhead{Fit} & \colhead{a} & \colhead{b}& \colhead{c} & \colhead{RMS}
}
\startdata
$\alpha$ & 0.057$\pm$0.064 & 0.690$\pm$0.108 & 0.581$\pm$0.190 & 0.10 \\
Mg       & 0.175$\pm$0.074 & 0.498$\pm$0.139 & 0.479$\pm$0.278 & 0.16 \\
Ca       & 0.075$\pm$0.070 & 0.639$\pm$0.125 & 0.532$\pm$0.216 & 0.12 \\
Ti       & 0.047$\pm$0.070 & 0.657$\pm$0.124 & 0.534$\pm$0.211 & 0.11 \\
Halo     & 0.237$\pm$0.013 & 0.229$\pm$0.032 & 0.270$\pm$0.021 & 0.19 \\
\enddata
\tablecomments{
The parameters for the $\alpha$, Mg, Ca, and Ti fits were derived using the full RRL+HB
sample. The Halo fit was derived considering all objects in the left panels of 
Fig.~\ref{fig:alpha_comparison_bulgehalodisks} with the exception of the RRL+HB sample.
}
\end{deluxetable}

The spread in $\alpha$-abundance steadily decreases when moving 
from the metal-poor/metal-intermediate into the more metal-rich ([Fe/H]$\ge$-1.0) regime. 
The Lit-HB sample does not reach higher metallicities, but this change in spread appears
even when considering only the RRLs. Indeed, the spread in alpha element abundance decreases 
from $\sim$0.5 dex to $\sim$0.2 dex. The position of this sharp decrease 
in $\alpha$-abundance, the so-called "knee", is traditionally interpreted as evidence of
the growing impact of SNe Ia. While the SNe II, with their short time scales, quickly
enrich the interstellar medium with mainly $\alpha$-elements and some iron, the SNe Ia,
with much longer time scales, begin to enrich the interstellar medium when it is already
at higher metallicities, injecting it with mostly iron and causing a quick decrease of the
$\alpha$-to-iron ratio. 

The RRL and HB stars appear to be, at fixed iron abundance, more alpha enhanced than typical
Halo objects (left panel of Fig.~\ref{fig:alpha_comparison_bulgehalodisks}) in the metal-poor 
([Fe/H]$\le$=-2.0) regime and more alpha poor than typical Halo objects in the metal-rich ([Fe/H]$\ge$-1.0)
 regime. Indeed, a fit with the same logarithmic form shown above but applied to these typical Halo objects 
is indicated by a dashed grey line in both panels of Fig.~\ref{fig:alpha_weighted_averages}. The
corresponding parameters are listed in Tab.~\ref{tab:fit_parameters}. This plain evidence could imply that the role played by SNe II in the 
Halo chemical enrichment was more 
crucial in the metal-poor than in the metal-intermediate/metal-rich regime.  

There is mounting evidence for a sizable sample of metal-rich HB stars that are 
also $\alpha$-poor \citep[see Figure 12 in][]{2018AJ....155..240A}. Indeed, for iron abundances 
larger than $\approx$-0.2 dex their $\alpha$-element abundance is either solar or lower. 
This finding together with our results based on RRLs indicates that the chemical 
enrichment in a significant fraction of metal-rich old field stars was mainly driven by SNe Ia with 
a minor contribution from SNe II. The lockstep variations we observed for [X/H] for the
three species, and in particular the spread in Mg that is comparible to that of the other species (Sect.\ref{ch:results_individual}), 
point towards an early chemical enrichment driven by a homogeneous initial mass function over a wide
range in iron abundance.

\begin{figure}
\includegraphics[width=\columnwidth]{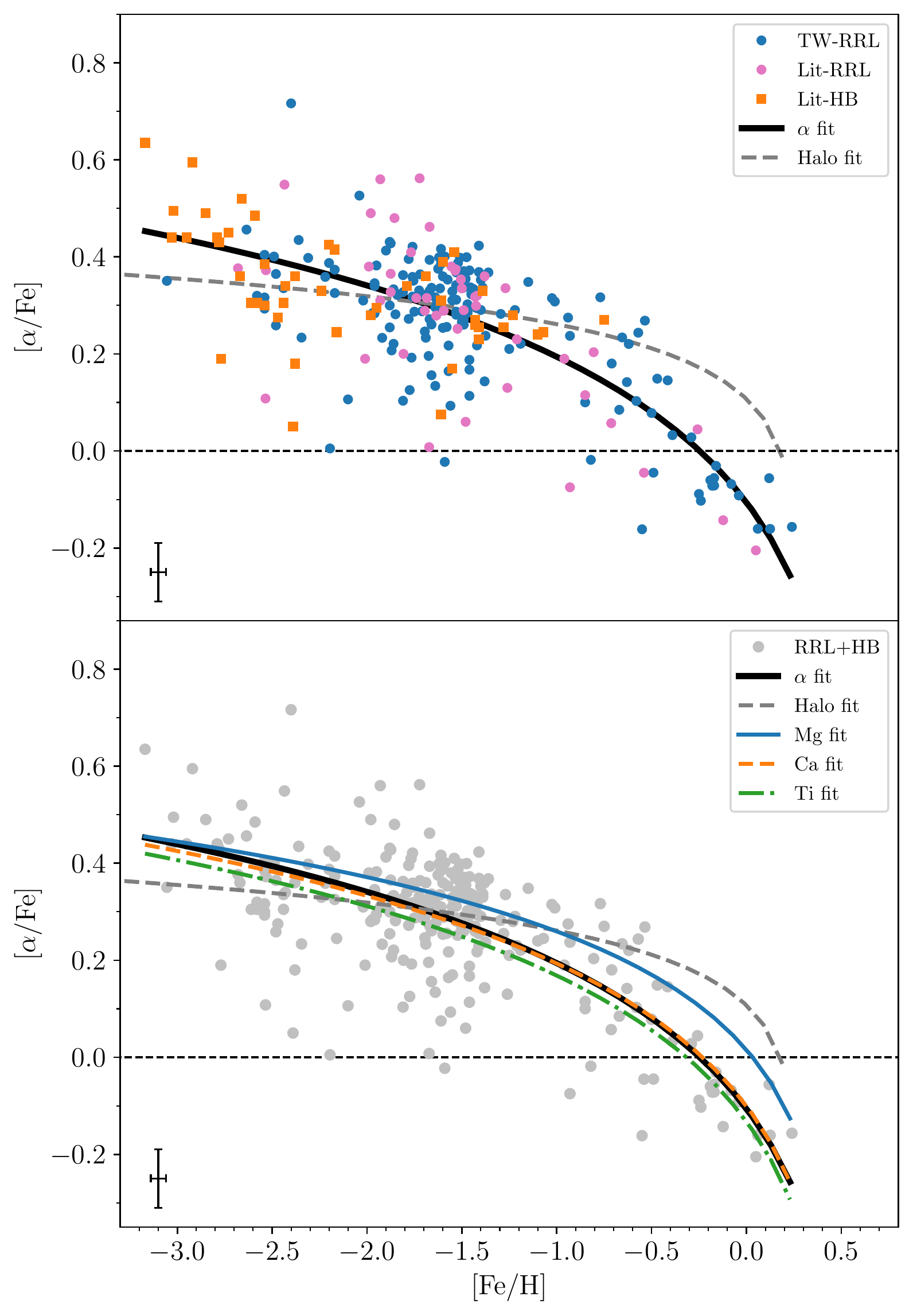}
\caption{
\emph{Top:} Average [$\alpha$/Fe] abundances using Mg, Ca, and Ti, for the stars in 
the TW-RRL (blue dots), Lit-RRL (pink dots), and Lit-HB (orange squares) samples. See text for 
details on the averaging method. The full black line shows the logarithmic fit for the RRL and 
Lit-HB samples, and the dashed grey line shows the logarithmic fit for the Halo field stars and 
GCs included the left panels of Fig.~\ref{fig:alpha_comparison_bulgehalodisks}.
\emph{Bottom:} Same as the top, but with the full RRL+HB sample shown as grey dots. The logarithmic
fits for the individual species are shown for Mg (blue solid line), Ca (orange dashed line), and
Ti (green dot-dashed line).
\label{fig:alpha_weighted_averages}}
\end{figure}


\section{Summary and future perspectives} \label{ch:final_remarks}
We performed the largest and most homogeneous measurement of $\alpha$-element (Mg, Ca, Ti) 
and iron abundances for field RRLs (162) by using high-resolution spectroscopy. 
This dataset was complemented with similar abundance estimates available in the 
literature for 46 field RRLs transformed into our metallicity scale by using 
objects in common. We ended up with a sample of old (t$\ge$ 10 Gyr) stellar 
tracers (208 RRLs: 169 RRab, 38 RRc, 1 RRd) covering more than three dex in iron 
abundance (-3.00$\le$[Fe/H]$\le$0.24). Note that the targets were selected to have 
Galactocentric distances ranging from $\sim$5 to $\sim$ 25 kpc. Therefore, they are solid beacons to 
investigate the early chemical enrichment of the Galactic Halo. 

We found that Mg, Ca, and Ti abundances vary, within the errors, in lock step with 
one another and have similar scatter over the entire range in iron abundance. 
Furthermore, the trend in the [$\alpha$/Fe] versus [Fe/H] plane displayed by field 
RRLs only partially follows the trend typical of other Halo stellar populations. 
RRLs in the metal-poor regime, appear to be systematically more $\alpha$-enhanced 
by $\sim$0.1 dex, while in the metal-rich regime they are more $\alpha$-poor 
by $\sim$0.3 dex, i.e. a factor of three larger than the typical uncertainties. 
This is the first time this depletion in $\alpha$-elements is detected on the 
basis of a large, homogeneous and coeval sample of old stellar tracers. 

A comparison with nearby classical dwarf galaxies and ultra-faint dwarf galaxies reveals a
remarkable agreement between the Halo RRL and RGs in the Sagittarius dSph
galaxy in the metal-rich regime. In the the metal-poor regime, beyond the range 
of the Sagittarius dSph sample, the RRL display a better agreement with the ultra-faint 
dwarf galaxies than with more massive dwarf galaxies.

To further constrain the role played by stellar age in the early chemical enrichment 
of the Halo, we took also into account similar elemental abundances for 46 field blue 
and red HB stars provided by \citet{2010AJ....140.1694F}. These stars are either slightly 
hotter (blue) or slightly cooler (red) than RRLs, however, they share the same 
evolutionary phase (central helium burning) and the same old (t$\ge$ 10 Gyr), 
low-mass progenitors. Theory and observations indicate that they only differ in their 
envelope mass. We found that RRLs and HB stars show the same
trends in the [$\alpha$/Fe] versus [Fe/H] planes. These findings support 
the Halo early chemical enrichment, here traced by unambiguously old stellar tracers.   

To overcome the possible occurrence of significant continuum placement uncertainty or
saturated lines, we carefully selected lines 
with equivalent widths between 15 and 150 m\AA{}. Moreover, we selected several lines 
that could be measured over a significant fraction of the range in iron covered by the 
current RRL sample in order to verify that no systematic differences between different
lines and chemical species  influenced the trend in the [$\alpha$/Fe] versus [Fe/H] plane. 
Finally, we also performed a comparison with spectroscopic standards (Arcturus) 
and with field metal-rich red HB stars \citep{2018AJ....155..240A}. We found 
that our iron and $\alpha$-element abundances are, within the errors, in remarkable 
agreement with similar estimates available in the literature.    

Chemical evolution models, for the chemical species discussed in this investigation, 
point to different dependencies of the yields on the stellar mass and the metallicity 
regime. This means that the current findings can be soundly adopted to constrain 
the chemical enrichment history of the Halo. In passing we also note that RRLs and HB 
stars cover a very narrow range in stellar masses, therefore the comparison with similar 
Halo stellar tracers can provide useful insights 
into the role played by the initial mass function and the star formation rate 
during the Halo early formation. 

Our findings concerning the impact that stellar age has on the analysis of the different 
[$\alpha$/Fe] vs [Fe/H] planes is very promising. New spectroscopic surveys 
(
WEAVE, \citealt{2016ASPC..507...97D};
4MOST, \citealt{2019Msngr.175....3D}; 
GALAH, \citealt{2015MNRAS.449.2604D};  
H3, \citealt{2019ApJ...887..237C}; 
SDSSV, \citealt{2017arXiv171103234K}) 
based on high resolution optical spectra will provide in a few years detailed 
Halo elemental abundances not only for blue HB and red HB stars, but also for RRLs. This means the unique 
opportunity to investigate the fine structure in time and in Galactocentric distance of the 
Halo early chemical enrichment. A similar quantitative jump is also planned for the chemical enrichment of the Galactic Bulge. 
Thanks to current 
(APOGEE, \citealt{2017AJ....154...94M}; 
WINERED, \citealt{2016SPIE.9908E..5ZI}) 
and near future 
(CRIRES+ at VLT, \citealt{2014SPIE.9147E..19F}; 
MOONS at VLT, \citealt{2014SPIE.9147E..0NC};
ERIS at VLT, \citealt{2018SPIE10702E..09D};
PFS at Subaru, \citealt{2018SPIE10702E..1CT}) 
spectroscopic surveys, detailed
elemental abundances will also become available. This means the opportunity to 
constrain on a quantitative basis the chemical enrichment and the timescale of the 
Galactic spheroid, i.e. both the Halo and the Bulge.       

The current spectroscopic measurements are a fundamental stepping stone for a detailed 
comparison between chemical evolution models and observations. Indeed, our RRL sample was built to
provide a clean (concerning the age distribution) and homogeneous 
(concerning the methodological approach and the spectroscopic data set) observational framework to compare 
with theoretical predictions \citep{2008A&A...481..691C,2011A&A...531A..72S,2018ApJS..237...13L}.

\acknowledgments

We thank the anonymous referee for his/her positive words regarding the content and cut of the paper, and for his/her pertinent suggestions that helped improve its readability. 
It is a pleasure to thank F. Matteucci and M. Palla for several detailed discussions concerning the role of 
$\alpha$-elements in constraining chemical evolution models, and E. Carretta for insightful discussions 
concerning the $\alpha$-elements in field and cluster stars.

This research has made use of the National Aeronautics and Space Administration (NASA) Astrophysics Data System,
the National Institute of Standards and Technology (NIST) Atomic Spectra Database, the JVO Portal\footnote{
\url{http://jvo.nao.ac.jp/portal/}} operated by ADC/NAOJ, and the ESO Science Archive Facility. 

Based on observations made with the Italian Telescopio Nazionale Galileo (TNG) operated on the island of La Palma 
by the Fundaci\'{o}n Galileo Galilei of the INAF (Istituto Nazionale di Astrofisica) at the Spanish Observatorio del 
Roque de los Muchachos of the Instituto de Astrofisica de Canarias.

Some of the observations reported in this paper were obtained with the Southern African Large Telescope (SALT).

Based on observations collected at the European Organisation for Astronomical Research in the Southern Hemisphere 
under ESO programmes 0100.D-0339, 0101.D-0697, 0102.D-0281, 076.B-0055, 077.B-0359, 077.D-0633, 079.A-9015, 079.D-0262, 
079.D-0462, 079.D-0567, 082.C-0617, 083.B-0281, 083.C-0244, 094.B-0409, 095.B-0744, 097.A-9032, 098.D-0230, 189.B-0925, 
267.C-5719, 297.D-5047, 67.D-0321, 67.D-0554, 69.C-0423, 71.C-0097, 0100.D-0273, 083.C-0244, 098.D-0230.

Funding for the SDSS and SDSS-II has been provided by the Alfred P. Sloan Foundation,
the Participating Institutions, the National Science Foundation, the U.S. Department of
Energy, the National Aeronautics and Space Administration, the Japanese Monbukagakusho,
the Max Planck Society, and the Higher Education Funding Council for England. The SDSS
Web Site is \url{http://www.sdss.org/.}
The SDSS is managed by the Astrophysical Research Consortium for the Participating
Institutions. The Participating Institutions are the American Museum of Natural History,
Astrophysical Institute Potsdam, University of Basel, University of Cambridge, Case Western
Reserve University, University of Chicago, Drexel University, Fermilab, the Institute for Advanced Study, 
the Japan Participation Group, Johns Hopkins University, the Joint Institute
for Nuclear Astrophysics, the Kavli Institute for Particle Astrophysics and Cosmology, the
Korean Scientist Group, the Chinese Academy of Sciences (LAMOST), Los Alamos National
Laboratory, the Max-Planck-Institute for Astronomy (MPIA), the Max-Planck-Institute for
Astrophysics (MPA), New Mexico State University, Ohio State University, University of
Pittsburgh, University of Portsmouth, Princeton University, the United States Naval Observatory, 
and the University of Washington.

We acknowledge financial support from US NSF under Grants AST-1714534 (MM, JPM) and AST1616040 (CS). 
EKG, BL, AJKH, ZP, and HL were supported by the Deutsche Forschungsgemeinschaft (DFG, German Research Foundation)
- Project-ID 138713538 - SFB 881 ("The Milky Way System", subprojects A03, A05, A11). EV acknowledges 
the Excellence Cluster ORIGINS Funded by the Deutsche Forschungsgemeinschaft (DFG, German Research Foundation) 
under Germany's Excellence Strategy \-- EXC \-- 2094 \--390783311. AJKH and ZP gratefully acknowledge funding 
by the Deutsche Forschungsgemeinschaft (DFG, German Research Foundation) -- Project-ID 138713538 -- SFB 881 
(``The Milky Way System''), subprojects A03, A05, A11.

\appendix

\section{Comparison of equivalent widths}  \label{ap:EW_comparisons}
As a simple sanity check, we compared the EWs among pairs of stars with similar effective temperatures and with either 
similar iron or $\alpha$-abundance (Fig.~\ref{fig:EWcompareL}). The number of such pairs is limited due to the need 
not only of similar T$_{\text{eff}}$ but also of a significant number of lines in common. The comparison is particularly 
difficult for stars with lower metallicities or at higher effective temperatures, as they have fewer lines and higher
uncertainties overall. 
 
The effective temperature is the parameter that most strongly affects the abundance of each individual line. Thus, a
comparison further supports the real variations of [Fe/H] and [$\alpha$/H] among our stars. For a difference of up to 
approximately 0.15 dex, the EWs for both stars visually coincide. This can be seen in the second row of the left column 
of Fig.~\ref{fig:EWcompareL}. For the other panels, the pair of stars have a difference in either [Fe/H] or [$\alpha$/H]
that can be visually detected by the fact that the EWs for the abundance that is similar show an identity relation,
while the EWs for abundance that is different are shifted from the identity to either higher or lower values.  

\begin{figure*}
\includegraphics[width=\columnwidth]{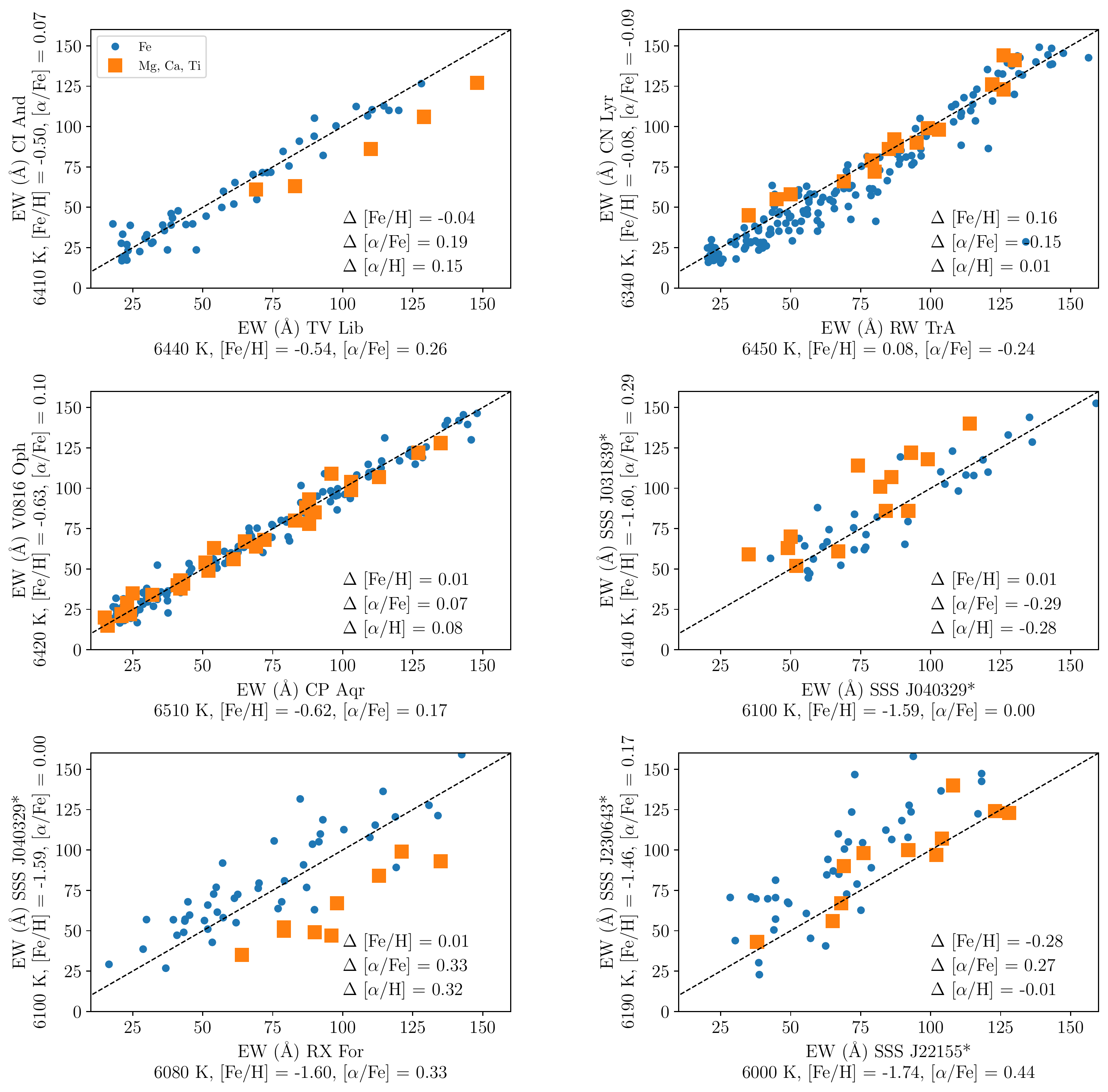}
\caption{Comparison of equivalent widths for iron (blue dots) and $\alpha$-element (orange
squares) lines for stars at similar effective temperatures. The atmospheric parameters of
interest for each star are shown in the axes. The difference between [Fe/H], [$\alpha$/H],
and [$\alpha$/Fe] are shown on the right left corner of each panel.
\label{fig:EWcompareL}}
\end{figure*}

\section{NLTE corrections and the trend of individual lines} \label{ap:NLTE_individuallines}
The absorption lines of a given chemical species display equivalent widths that depend on the 
abundance of the element in question, which is directly related to the overall metallicity [M/H] and, consequently,
to the [Fe/H] as well. In other words, the EWs of a given element will be smaller at lower metallicities, and 
increase with increasing metallicity. This is of particular importance when analysing the [X/Fe] versus [Fe/H] trend 
because a given line may be too weak at low [Fe/H] or too strong at high [Fe/H], and therefore only 
measured in a limited range of [Fe/H] values. Using several lines of different strengths ensures that the whole 
metallicity range is covered, but the transition parameters of different lines are subjected to uncertainties that 
may result in significant disagreements between them and create spurious trends if one line is only available at
lower metallicities, and the other line only at higher metallicities. Therefore, the presence of one single line
that covers a wide metallicity range is extremely valuable in order to confirm that any trends are real and not
due to such systematics.

A few of the lines adopted in this work cover the entire metallicity range of the current sample. For Mg, no line is 
present in all metallicity regimes, but the domains of different lines are superposed in such a way that it is possible 
to verify they are in the same scale. Moreover, the overall behavior of Mg is in agreement with that of Ca and Ti. This 
is shown in Fig.~\ref{fig:lines_metallicity_range}. 

Different lines in the same chemical species may be subjected to different levels of NLTE effects. We obtained the values 
of NLTE corrections using the MPIA NLTE Spectrum Tools\footnote{Available at \url{http://nlte.mpia.de/}}. The corrections
for Mg \citep{2017ApJ...847...15B}, Ca \citep{2007A&A...461..261M}, and Ti \citep{2011MNRAS.413.2184B} lines were computed
using 1D plane-parallel models for a set of typical RRL atmospheric parameters in the whole metallicity range covered by 
our sample. A few lines adopted in this work were not available for this analysis, namely the Mg lines at 8712.69 and 
8717.83 \AA{}, the Ca lines at 5581.97, 5601.29, 6471.66, 6493.78, 6499.65, and 6717.69 \AA{}, the TiI lines at 3729.81, 3741.06, 
5036.46, and 5038.4 \AA{}, and the TiII lines at 4464.45, 6606.96, and 7214.73 \AA{}. The results for the remaining lines are as 
follows.

i) {\em Mg}: Most corrections for the seven available lines were under $\pm$0.05 dex, with the exception of the metal-poor
tail, where four lines had corrections of the order of +0.14.

ii) {\em Ca}: Most lines had corrections of the order of a staggering +0.4 dex. However, three lines (6166.4, 6449.8, and 
6455.6 \AA{}) display vanishing corrections and yet, in our results, these lines are in good agreement with all others in any 
given star where they appear. We also note that choosing the spherical 1D models produced corrections as high as 1.0 dex. 
The source paper for the corrections does not quite cover the atmospheric parameters of RRL, but the closest values provide
corrections of the order of 0.2 dex in the metal-poor regime for some of the lines, with smaller positive or negative 
corrections for the metal-rich regime. We have included the 6455.6 \AA{} line as goldenrod stars in Fig.~\ref{fig:lines_metallicity_range}.
If the corrections were adopted, this line would remain largely unchanged, while the others were shifted +0.4 dex. 

iii) {\em TiI}: All available lines displayed corrections of the order of +0.15 dex in the metal-rich regime, increasing to about 
+0.30 dex in the metal-poor regime. This would further increase the slope in the [TiI/Fe] versus [Fe/H] plane. Interestingly,
the results for TiI without any NLTE have a tighter scatter than those for TiII with or without NLTE considerations.

iv) {\em TiII}: All available corrections were vanishing, except for a few lines where a shift of the order of +0.12 dex was
present in the metal-poor regime. We note, however, that the corrections for the line at 4911.19 \AA{}, shown as goldenrod
stars in 
Fig.~\ref{fig:lines_metallicity_range}, were zero.

Without considering the NLTE corrections, a few lines in our sample seem to be systematically higher than others (see e.g. 
the TiII lines at 4805.09 and 5185.90 \AA{} in Fig.~\ref{fig:lines_metallicity_range}), however other lines follow either of 
the two sequences set by these lines, or remain between them. As the trend in [X/Fe] versus [Fe/H] is the same for both
sequences, taking the average value among all lines preserves it, and the higher $\sigma$ of each $\alpha$-element 
accounts for this decision to include all available lines. The LTE analysis preserves both the internal consistency of our
investigation, and the possibility to readily compare it against other data sets in the literature.

\begin{figure}
\begin{centering}
\includegraphics[width=0.5\columnwidth]{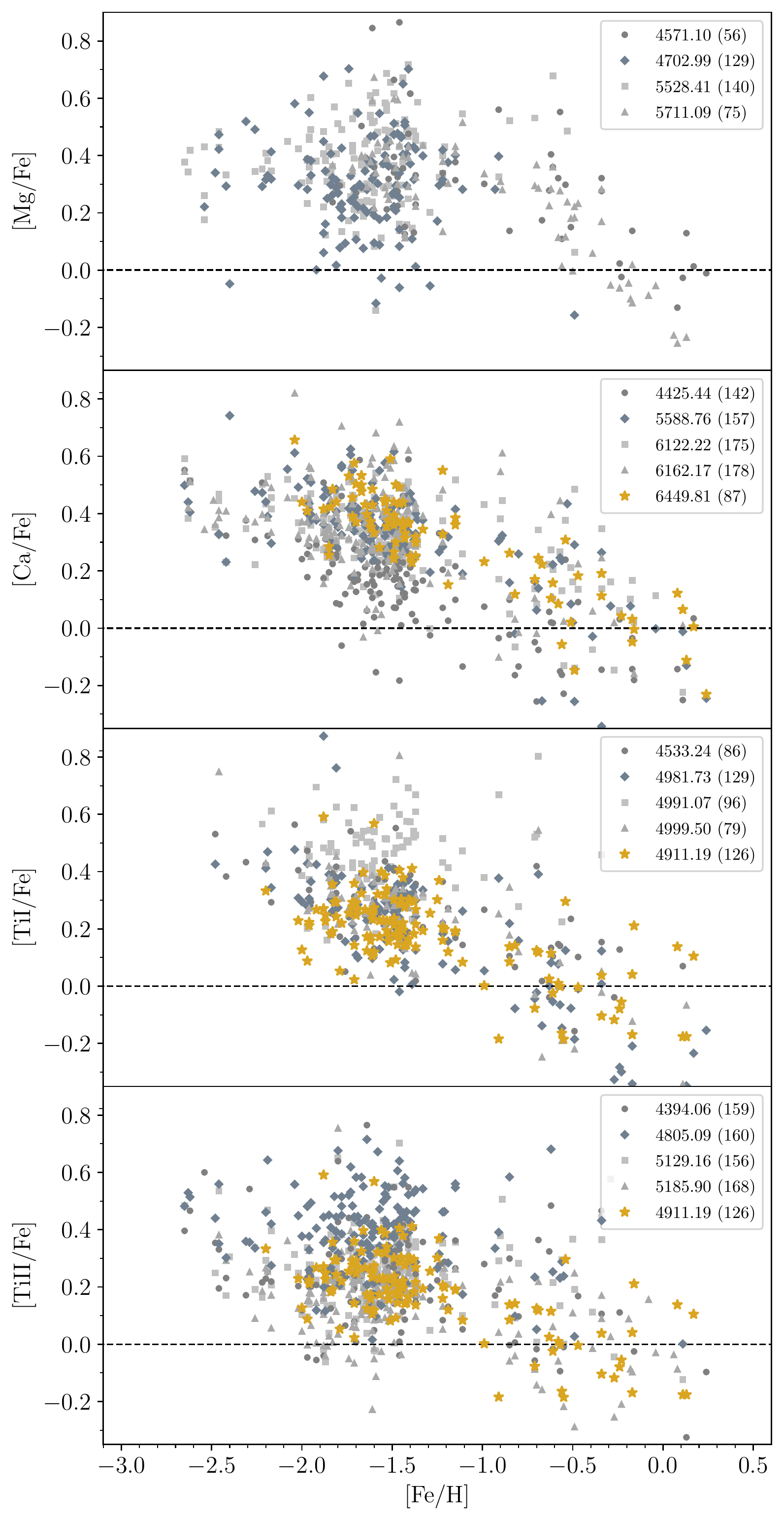}
\caption{The [X/Fe] versus [Fe/H] trend for individual lines for the three chemical species considered
in this work. The lines shown as grey symbols are those with the greatest number of measurments. In red,
we show two lines that do not require any NLTE correction. See text for details.
\label{fig:lines_metallicity_range}}
\end{centering}
\end{figure}


\bibliography{bibliography}{}
\bibliographystyle{aasjournal}

\end{document}